\documentclass[11pt,twoside]{article} 
\usepackage{times,fancyhdr}
\usepackage[dvips]{graphicx}

\begin{document}
\title{
{\begin{flushleft}
\vskip 0.45in
{\normalsize\bfseries\textit{Chapter~1}}
\end{flushleft}
\vskip 0.45in \bfseries\scshape Nuclear limits on properties of
pulsars and gravitational waves}}
\author{\bfseries\itshape Plamen G. Krastev\thanks{E-mail address: pkrastev@sciences.sdsu.edu}\\
Department of Physics and Astronomy, Texas A\&M University-Commerce, \\
P.O. Box 3011, Commerce, TX 75429, U.S.A.\\
Department of Physics, San Diego State University, \\
5500 Campanile Drive, San Diego CA 92182-1233, U.S.A.
\\\\
\bfseries\itshape Bao-An Li\thanks{E-mail address: Bao-An\_Li@tamu-commerce.edu}\\
Department of Physics and Astronomy, Texas A\&M University-Commerce,\\
 P.O. Box 3011, Commerce, TX 75429, U.S.A.}
\date{\today}
\maketitle \thispagestyle{empty} \setcounter{page}{1}
\thispagestyle{fancy} \fancyhead{} \fancyhead[L]{Pulsars: Theory,
Categories and Applications,\\
Editor: Frank Columbus et al., pp. {\thepage-\pageref{lastpage-01}}} 
\fancyhead[R]{ISBN 0000000000  \\
\copyright~2010 Nova Science Publishers, Inc.} \fancyfoot{}
\renewcommand{\headrulewidth}{0pt}

\vspace{2in}


\newpage

\pagestyle{fancy} \fancyhead{} \fancyhead[EC]{Plamen G. Krastev
and Bao-An Li} \fancyhead[EL,OR]{\thepage} \fancyhead[OC]{Nuclear
limits on properties of pulsars and gravitational waves}
\fancyfoot{}
\renewcommand\headrulewidth{0.5pt}

\begin{abstract}
Pulsars are among the most mysterious astrophysical objects in the
Universe and are believed to be rotating neutron stars formed in
supernova explosions. They are unique testing grounds of dense
matter theories and gravitational physics and also provide links
among nuclear physics, particle physics and General Relativity.
Neutron stars may exhibit some of the most extreme and exotic
characteristics that could not be found elsewhere in the Universe.
Their properties are largely determined by the equation of state
(EOS) of neutron-rich matter, which is the chief ingredient in
calculating neutron star structure and properties of related
phenomena, such as gravitational wave emission from deformed
pulsars. Presently, the EOS of neutron-rich matter is still very
uncertain mainly due to the poorly known density dependence of the
nuclear symmetry energy especially at supra-saturation densities.
Nevertheless, significant progress has been made recently in
constraining the density dependence of the nuclear symmetry energy
mostly at sub-saturation densities using terrestrial nuclear
reactions. While there are still some uncertainties especially at
supra-saturation densities, these constraints could provide useful
information on the limits of the global properties of pulsars and
the gravitational waves to be expected from them. Here we review
our recent work on constraining properties of pulsars and
gravitational radiation with data from terrestrial nuclear
laboratories.
\end{abstract}

\tableofcontents

\section{Introduction}

Pulsars exhibit a large array of extreme characteristics. They are
generally accepted to be rotating neutron stars -- the smallest
and densest stars known to exist. Matter in their cores is
compressed to huge densities ranging from the density of normal
nuclear matter, $\rho_0\approx 0.16fm^{-1}$, to an order of
magnitude higher~\cite{Glendenning:2000a}. The number of baryons
forming a neutron star is in the order of $A\approx{10^{57}}$.
Understanding properties of matter under such extreme conditions
of density (and pressure) is still far from complete and
represents one of the most important but also challenging problems
in modern physics.

Because of their strong gravitational binding neutron stars can
rotate very fast~\cite{Bejger:2006hn}. The first millisecond
pulsar PSR1937+214, spinning at $\nu=642 Hz$~\cite{Backer:1982},
was discovered in 1982, and during the next decade or so almost
every year a new one was reported. In the recent years the
situation changed considerably with the discovery of an
anomalously large population of millisecond pulsars in globular
clusters~\cite{Weber:1999a}, where the density of stars is roughly
1000 times that in the field of the galaxy and which are therefore
very favorable sites for formation of rapidly rotating neutron
stars which have been spun up by the means of mass accretion from
a binary companion. Presently, the number of the observed pulsars
is close to 2000, and the detection rate is rather high.

In 2006 Hessels et al.~\cite{Hessels:2006ze} reported the discovery
of a very rapid pulsar J1748-2446ad, rotating at $\nu=716Hz$ and
thus breaking the previous record (of $642Hz$). Rapid rotation could
affect significantly the neutron star structure and properties,
especially if the rotational frequency is near the Kepler frequency
for the star. (The Kepler, or mass-shedding frequency, is the
highest possible frequency an object bound by gravity can have
before it starts to loose mass at the equator.) Pulsars with masses
above $1M_{\odot}$ enter the rapid-rotation regime, if their
rotational frequencies are higher than 1000Hz~\cite{Bejger:2006hn}.
Although, presently no pulsar is confirmed to rotate above the
1000Hz limit, theory does not exclude the possibility for existence
of such rapidly rotating neutron stars and therefore it is important
to predict their properties. (Here we should mention the recent
discovery of X-ray burst oscillations from the X-ray transient XTE
J1739–285~\cite{Kaaret:2006gr}, which could suggest that it contains
an extremely rapidly rotating neutron star spinning at 1122 Hz.
However, this observation has not been confirmed yet.) Neutron stars
are natural astrophysical laboratories of dense
matter~\cite{Weber:1999a}. Rotating neutron stars appear as much
better probes for the structure of dense baryonic matter than static
ones, primarily because the particle compositions in rotating
neutron stars are not frozen in, as it is the case for non-rotating
neutron stars, but are varying with time~\cite{Weber:2009yd}. The
associated density changes could be as large as
60\%~\cite{Weber:1999a} in neutron stars in binary stellar systems
(e.g., low-mass X-ray binaries), which are being spun up to higher
rotational frequencies, or isolated rotating neutron stars (e.g.
isolated millisecond pulsars) which are spinning down to lower
frequencies because of the gravitational radiation, electromagnetic
dipole radiation, and a wind of electron-positron
pairs~\cite{Weber:2009yd}.

Since neutron stars are objects of extremely condensed matter, the
geometry of space-time is considerably altered from that of a flat
space. Therefore, the construction of realistic models of neutron
stars has to be done in the framework of General
Relativity~\cite{Glendenning:2000a,Weber:1999a}. Detailed
knowledge of the equation of state (EOS) of stellar matter over a
very wide range of densities is required for solving the
neutron-star structure equations. At present time the behavior of
matter under extreme densities such as those found in the
interiors of neutron stars is still highly uncertain and relies
upon, often, rather controversial theoretical predictions. On the
other hand, fortunately, heavy-ion reactions provide unique means
to constrain the EOS of dense nuclear matter in terrestrial
laboratories~\cite{Danielewicz:2002pu}. One of the major
uncertainties of the EOS is the density dependence of the nuclear
symmetry
energy~\cite{Lattimer:2004pg,Steiner:2004fi,Krastev:2006ii},
$E_{sym}(\rho)$, which is the difference between the nucleon
specific energies in pure neutron matter and symmetric nuclear
matter. Due to its importance for the neutron star structure and
many other ramifications in astrophysics and cosmology,
determining the density dependence of the nuclear symmetry energy
has been a major thrust of research for the intermediate energy
heavy-ion community. Although extracting information about the
density dependence of the nuclear symmetry energy from heavy-ion
reactions is not an easy task due to the complicated roles of the
isospin degree of freedom in the reaction dynamics, several
promising probes of the symmetry energy have been suggested and
tested in recent
years\cite{Li:1997rc,Li:2000bj,Li:2002qx,Li:1997px,LCK08} (see
also Refs.~\cite{Danielewicz:2002pu,Li:2001a,Baran:2004ih} for
reviews). In fact, some significant progress has been made in
determining the $E_{sym}(\rho)$ at subsaturation densities using:
(1) isospin diffusion~\cite{Tsang:2004} and
isoscaling~\cite{Tsang:2001} in heavy-ion reactions at
intermediate
energies~\cite{Shi:2003np,Chen:2005a,Steiner:2005rd,Li:2005jy,tsa},
(2) sizes of neutron skins in heavy
nuclei~\cite{Steiner:2004fi,Horowitz:2000xj,Horowitz:2002mb,ToddRutel:2005fa},
(3) the Pygmy dipole resonance~\cite{Cen09} and (4) masses of
nuclei~\cite{Dan09}. At supranormal densities, a number of
potential probes of the symmetry energy have been
proposed~\cite{LCK08,Chen:2007} although the available data has
been very limited so far. Interestingly, circumstantial evidence
for a super-soft $E_{sym}(\rho)$ at supra-saturation densities was
reported recently \cite{Xiao09} based on the analysis of the
FOPI/GSI data on pion production~\cite{Rei07}. It is also very
encouraging to notice that several dedicated experiments are being
planned at several laboratories to study in more detail the
symmetry energy at supra-saturation densities using high energy
heavy-ion reactions induced by both stable and radioactive beams.

While global properties of spherically symmetric static neutron
stars have been studied extensively and comprehensive literature
exists, e.g. Refs.
\cite{Lattimer:2004pg,Steiner:2004fi,Krastev:2006ii,Prakash:2001rx,Lattimer:2000kb,Heiselberg:1999mq,Heiselberg:2000dn,Yakovlev:2004iq},
properties of (rapidly) rotating neutron stars have been
investigated to lesser extent. Models of (rapidly) rotating neutron
stars have been constructed by several research groups with various
degree of approximation
\cite{Weber:1999a,Hartle:1967he,Hartle:1968si,Friedman:1986tx,Bombaci:2000rc,1990ApJ...355..241L,1989MNRAS.237..355K,1994ApJ...424..823C,Stergioulas:1994ea,Stergioulas:1997ja,1998PhRvD..58j4020B,1993A&A...278..421B,2002A&A...381L..49A}
(see Ref.~\cite{Stergioulas:2003yp} for a review). (Rapidly)
rotating, elliptically deformed neutron stars could be one of the
major candidates for emitting gravitational waves (GWs) -- tiny
ripples in space-time predicted by the theory of General
Relativity~\cite{Einstein:1918}. Although GWs have not been detected
directly yet, indirect evidence do exist~\cite{WT:2004}. GWs are
characterized by a small dimensionless strain amplitude, $h_0$,
which, in addition to the pulsar's distance to detector, depends on
the neutron star structure determined by the underlying EOS of
stellar matter. Thus the EOS is instrumental for predicting the
strength of gravitational radiation emitted from neutron stars.

The partially constrained EOS by the available terrestrial nuclear
laboratory data has been used in studying several properties of
neutron stars, including the mass-radius
correlation~\cite{Li:2005sr}, the surface temperature of neutron
stars in connection with the changing rate of the gravitational
constant $G$~\cite{Kra07}, the core-crust transition
density~\cite{Xu09a}, the frequency and damping time of the w-mode
of gravitational waves~\cite{DHW09a} and the gravitational binding
energy of neutron stars~\cite{Newton09}. In this chapter we
concentrate on reviewing several selected properties of fast
pulsars and the gravitational waves expected from them mostly
based on our recent work reported in
Refs.~\cite{KLW2,WKL:2008ApJ,Krastev:2008PLB}.

\section{The equation of state of neutron-rich nuclear matter
partially constrained by recent data from terrestrial heavy-ion
reactions}

In this section, we first outline the theoretical tools one uses
to extract information about the EOS of neutron-rich nuclear
matter from heavy-ion collisions. We put the special emphasis on
exploring the density-dependence of the symmetry energy as the
study on the EOS of symmetric nuclear matter with heavy-ion
reactions is better known to the astrophysical community and it
has been extensively reviewed, see e.g.,
Refs.~\cite{Danielewicz:2002pu,Steiner:2004fi,Lattimer:2007} for
recent reviews. We will then summarize the latest constraints on
the density dependence of the symmetry energy at sub-saturation
densities extracted from heavy-ion reactions at intermediate
energies. Finally, we address the question of what kind of
isospin-asymmetry, especially for dense matter, can be reached in
heavy-ion reactions.

Heavy-ion reactions provide unique means to create dense nuclear
matter in terrestrial laboratories similar to those found in the
core of neutron stars. While unlike the matter in neutron stars the
dense matter created in heavy-ion reactions is hot, accurate
information about the EOS of cold matter can be extracted reliably
by modeling carefully the kinetic part of the EOS during heavy-ion
reactions. Depending on the beam energy, impact parameter and the
reaction system, various hadrons and/or partons may be created
during the reaction. To extract information about the EOS of dense
matter from heavy-ion reactions requires careful modeling of the
reaction dynamics and selection of sensitive observables. Among the
available tools, computer simulations based on the
Boltzmann-Uehling-Uhlenbeck ({\sc buu}) transport theory have been
very useful, see, e.g., Refs.~\cite{Danielewicz:2002pu,Bertsch:1988}
for reviews. The evolution of the phase space distribution function
$f_i(\vec{r},\vec{p},t)$ of nucleon $i$ is governed by both the mean
field potential $U$ and the collision integral $I_{collision}$ via
the {\sc buu} equation
\begin{equation}
\frac {\partial f_i}{\partial t} + {\vec \nabla}_p U \cdot {\vec
\nabla}_r f_i - {\vec \nabla}_r U \cdot {\vec \nabla}_p f_i =
I_{collision}.
\end{equation}
Normally, effects of the collision integral $I_{collision}$ via both
elastic and inelastic channels including particle productions, such
as pions, are modeled via Monte Carlo sampling using either
free-space experimental data or calculated in-medium cross sections
for the elementary hadron-hadron scattering~\cite{Bertsch:1988}. The
collision integral is critical for modeling the kinetic part of the
EOS. Information about the EOS of cold nuclear matter is obtained
from the underlying mean-field potential U which is an input to the
transport model. By comparing experimental data on some carefully
selected observables with transport model predictions using
different mean-field potentials corresponding to various EOSs, one
can then constrain the corresponding EOS. The specific constrains on
the density dependence of the nuclear symmetry energy that we are
using in this work were obtained by analyzing the isospin diffusion
data~\cite{Tsang:2004} within the IBUU04 version of an isospin and
momentum dependent transport model~\cite{IBUU04}. In this model, an
isospin and momentum-dependent interaction (MDI)~\cite{Das:2002fr}
is used. With this interaction, the potential energy density $V(\rho
,T,\delta )$ at total density $\rho $, temperature $T$ and isospin
asymmetry $\delta$ is
\begin{eqnarray}
V(\rho ,T,\delta ) &=&\frac{A_{u}\rho _{n}\rho _{p}}{\rho _{0}}+\frac{A_{l}}{%
2\rho _{0}}(\rho _{n}^{2}+\rho _{p}^{2})+\frac{B}{\sigma
+1}\frac{\rho ^{\sigma +1}}{\rho _{0}^{\sigma }}(1-x\delta ^{2}) \nonumber \\
&+&\sum_{\tau ,\tau ^{\prime }}\frac{C_{\tau ,\tau ^{\prime
}}}{\rho_0}
\int \int d^{3}pd^{3}p^{\prime }\frac{f_{\tau }(\vec{r},\vec{p}%
)f_{\tau ^{\prime }}(\vec{r},\vec{p}^{\prime
})}{1+(\vec{p}-\vec{p}^{\prime })^{2}/\Lambda ^{2}}\label{MDIV}
\end{eqnarray}
In the mean field approximation, Eq. (\ref{MDIV}) leads to the
following single particle potential for a nucleon with momentum $\vec{p}$ and isospin $%
\tau $
\begin{eqnarray}
U_{\tau }(\rho ,T,\delta ,\vec{p},x)&=&A_{u}(x)\frac{\rho _{-\tau
}}{\rho_{0}}+A_{l}(x)\frac{\rho _{\tau }}{\rho _{0}}+B\left(
\frac{\rho}{\rho _{0}}\right) ^{\sigma }(1-x\delta ^{2})\nonumber\\
&-&8\tau x \frac{B}{\sigma +1}\frac{\rho ^{\sigma
-1}}{\rho_{0}^{\sigma}}\delta \rho_{-\tau }+\sum_{t=\tau ,-\tau }\frac{2C_{\tau ,t}}{\rho _{0}}\int d^{3}\vec{p}%
^{\prime
}\frac{f_{t}(\vec{r},\vec{p}^{\prime})}{1+(\vec{p}-\vec{p}^{\prime
})^{2}/\Lambda ^{2}},\label{MDIU}
\end{eqnarray}
where $\tau =1/2$ ($-1/2$) for neutrons (protons), $x$, $A_{u}(x)$,
$A_{\ell }(x)$, $B$, $C_{\tau ,\tau }$,$C_{\tau ,-\tau }$, $\sigma
$, and $\Lambda $ are all parameters given in Ref.
\cite{Das:2002fr}. The last two terms in Eq. (\ref{MDIU}) contain
the momentum dependence of the single-particle potential, including
that of the symmetry potential if one allows for different
interaction strength parameters $C_{\tau ,-\tau }$ and $C_{\tau,\tau
}$ for a nucleon of isospin $\tau $ interacting, respectively, with
unlike and like nucleons in the background fields. It is worth
mentioning that the nucleon isoscalar potential estimated from
$U_{isoscalar}\approx (U_{n}+U_{p})/2$ agrees with the prediction of
variational many-body calculations for symmetric nuclear
matter~\cite{wiringa} in a broad density and momentum
range~\cite{IBUU04}. Moreover, the EOS of symmetric nuclear matter
for this interaction is consistent with that extracted from the
available data on collective flow and particle production in
relativistic heavy-ion collisions up to five times the normal
nuclear matter~\cite{Danielewicz:2002pu}. On the other hand, the
corresponding isovector (symmetry) potential can be estimated from
$U_{sym}\approx (U_{n}-U_{p})/2\delta $. At normal nuclear matter
density, the MDI symmetry potential agrees very well with the Lane
potential extracted from nucleon-nucleus and (n,p) charge exchange
reactions available for nucleon kinetic energies up to about $100$
MeV~\cite{IBUU04}. At abnormal densities and higher nucleon
energies, however, there is no experimental constrain on the
symmetry potential available at present.

The different $x$ values in the MDI interaction are introduced to
vary the density dependence of the nuclear symmetry energy while
keeping other properties of the nuclear equation of state fixed.
Specifically, choosing the incompressibility $K_{0}$ of cold
symmetric nuclear matter at saturation density $\rho _{0}$ to be
$211$ MeV leads to the dependence of the parameters $A_{u}$ and
$A_{l}$ on the $x$ parameter according to
\begin{eqnarray}
A_{u}(x)=-95.98-x\frac{2B}{\sigma
+1},~A_{l}(x)=-120.57+x\frac{2B}{\sigma +1},
\end{eqnarray}
with $B=106.35~{\rm MeV}$.

\begin{figure}[t!]
\centering
\includegraphics[scale=0.7]{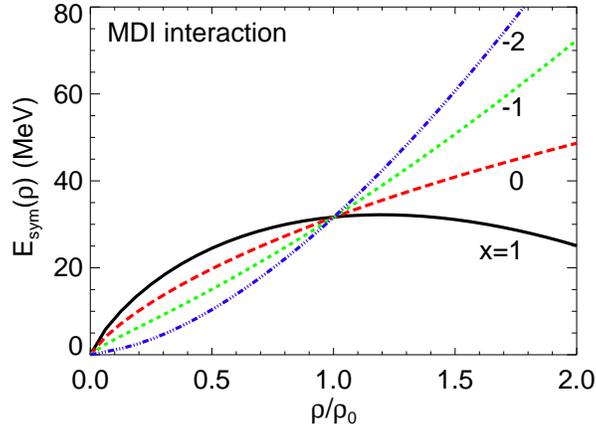}
\caption{The density dependence of the nuclear symmetry energy for
different values of the parameter $x$ in the MDI interaction. Taken
from \cite{Li05}.} \label{MDIsymE}
\end{figure}

With the potential contribution from Eq.~(\ref{MDIV}) and the
well-known contribution from nucleon kinetic energies in the Fermi
gas model, the EOS and the symmetry energy at zero temperature can
be easily obtained. As shown in Fig.~(\ref{MDIsymE}), varying the
parameter $x$ leads to a broad range of the density dependence of
the nuclear symmetry energy, similar to those predicted by various
microscopic and/or phenomenological many-body theories. As
previously demonstrated in~\cite{Li:2005jy} and~\cite{Li:2005sr},
only equations of state with $x$ between -1 and 0 have symmetry
energies in the sub-saturation density region consistent with the
isospin diffusion data and the available measurements of the skin
thickness of $^{208}Pb$ using hadronic probes. More recently
however, it was shown that at supra-saturation densities the
symmetry energy could be much softer than the extrapolation of the
symmetry energy with $x=0$~\cite{Xiao09}. The possibility for a
super-soft symmetry energy at supra-saturation densities could
have important consequences for the neutron star stability and the
nature of gravity at sub-millimeter distances~\cite{Wen:2009av}.
Nevertheless, as the first and a conservative step in our studies
of the properties of (rotating) neutron stars
\cite{KLW2,WKL:2008ApJ} and the gravitational waves expected from
them \cite{Krastev:2008PLB}, we have used $x=-1$ and $x=0$ to
compute the range of possible (rotating) neutron star
configurations. The $E_{sym}(\rho)$ with $x=0$ is also consistent
with the RMF prediction using the FSUGold
interaction~\cite{Piek07}. In this review, we thus consider only
the two limiting cases with $x=0$ and $x=-1$ as boundaries of the
symmetry energy consistent with the available terrestrial nuclear
laboratory data at sub-saturation densities.

\begin{table}[b!]
\caption{{\protect\small The parameters }$F${\protect\small \ (MeV),
}$G$
{\protect\small , }$K_{sym}${\protect\small \ (MeV), } $L$%
{\protect\small \ (MeV), and }$K_{asy}${\protect\small \ (MeV) for
different values of} $x${\protect\small. Taken from
\cite{Chen:2005a}.}}
\begin{center}
\begin{tabular}{ccccccc}\label{tab.1}
$x$ & \quad $F$ & $G$ & $K_{sym}$ & $L$ & $K_{asy}$ &  \\
\hline\hline
$1$  & $107.232$ & $1.246$ & $-270.4$ & $16.4$  & -368.8 &  \\
$0$  & $129.981$ & $1.059$ & $-88.6$  & $62.1$  & -461.2 &  \\
$-1$ & $3.673$   & $1.569$ & $94.1$   & $107.4$ & -550.3 &  \\
$-2$ & $-38.395$ & $1.416$ & $276.3$  & $153.0$ & -641.7 &  \\
\hline
\end{tabular}
\end{center}
\end{table}

To facilitate comparisons with other models in the literature, it is
useful to parameterize the $E_{sym}(\rho )$ from the MDI interaction
and list its characteristics. Within phenomenological models it is
customary to separate the symmetry energy into the kinetic and
potential parts, see, e.g.~\cite{Pra88b},
\begin{equation}
E_{sym}(\rho )=(2^{2/3}-1)\frac{3}{5}E_{F}^{0}(\rho /\rho
_{0})^{2/3}+E_{sym}^{\mathrm{pot}}(\rho).
\end{equation}
With the MDI interaction, the potential part of the nuclear symmetry
energy can be well parameterized by
\begin{equation}
E_{sym}^{\mathrm{pot}}(\rho ) =F(x)\rho /\rho_{0} +(18.6-F(x))(\rho
/\rho _{0})^{G(x)},
\end{equation}
with $F(x)$ and $G(x)$ given in Table \ref{tab.1} for $x=1$, $0$,
$-1$ and $-2$. The MDI parameterizations for the
$E_{sym}^{\mathrm{pot}}(\rho)$ are similar to but significantly
different from those used in Ref.~\cite{Pra88b}. Also shown in Table
\ref{tab.1} are other characteristics of the symmetry energy,
including its slope parameter $L$ and curvature parameter $K_{sym}$
at $\rho_0$, as well as the isospin-dependent part
$K_{\mathrm{asy}}$ of the isobaric incompressibility of asymmetric
nuclear matter~\cite{Chen:2005a}. The symmetry energy in the
subsaturation density region with x=0 and -1 can be roughly
approximately by $E_{sym}(\rho)\approx 31.6 (\rho/\rho_0)^{0.69}$
and $E_{sym}(\rho)\approx 31.6 (\rho/\rho_0)^{1.05}$, respectively.

What is the maximum isospin asymmetry reached, especially in the
supra-normal density regions, in typical heavy-ion reactions? How
does it depend on the symmetry energy? Do both the density and
isospin asymmetry reached have to be high simultaneously in order to
probe the symmetry energy at supra-normal densities with heavy-ion
reactions? The answers to these questions are important for us to
better understand the advantages and limitations of using heavy-ion
reactions to probe the EOS of neutron-rich nuclear matter and
properly evaluate their impacts on astrophysics.

\begin{figure}[t!]
\centering
\includegraphics[scale=0.4]{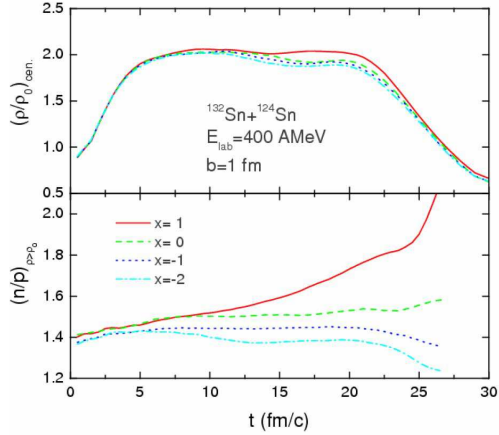}
\caption{Central baryon density (upper panel) and isospin asymmetry
(lower panel) of high density region in the reaction of $^{132}{\rm
Sn}+^{124}{\rm Sn}$ at a beam energy of 400 MeV/nucleon and an
impact parameter of 1 fm. Taken from Ref. \protect\cite{Li05}.}
\label{CentDen}
\end{figure}

To answer these questions we first show in Fig.~\ref{CentDen} the
central baryon density (upper window) and the average
$(n/p)_{\rho\geq \rho_0}$ ratio (lower window) of all regions with
baryon densities {\it higher than $\rho_0$} in the reaction of
$^{132}Sn+^{124}Sn$ at a beam energy of 400 MeV/nucleon and an
impact parameter of 1 fm. It is seen that the maximum baryon density
is about 2 times the normal nuclear matter density. Moreover, the
compression is rather insensitive to the symmetry energy because the
latter is relatively small compared to the EOS of symmetric nuclear
matter around this density. The high density phase lasts for about
15 fm/c from 5 to 20 fm/c for this reaction. It is interesting to
see in the lower window that the isospin asymmetry of the high
density region is quite sensitive to the density dependence of the
symmetry energy used in the calculation. The soft (e.g., $x=1$)
symmetry energy leads to a significantly higher value of
$(n/p)_{\rho\geq \rho_0}$ than the stiff one (e.g., $x=-2$). This is
consistent with the well-known isospin fractionation phenomenon in
asymmetric nuclear matter~\cite{Mul95,LiKo}. Because of the
$E_{sym}(\rho)\delta^2$ term in the EOS of asymmetric nuclear
matter, it is energetically more favorable to have a higher isospin
asymmetry $\delta$ in the high density region with a softer symmetry
energy functional $E_{sym}(\rho)$. In the supra-normal density
region, as shown in Fig.~\ref{MDIsymE}, the symmetry energy changes
from being soft to stiff when the parameter $x$ varies from 1 to -2.
Thus the value of $(n/p)_{\rho\ge \rho_0}$ becomes lower as the
parameter $x$ changes from 1 to -2. It is worth mentioning that the
initial value of the quantity $(n/p)_{\rho\ge \rho_0}$ is about 1.4
which is less than the average n/p ratio of 1.56 of the reaction
system. This is because of the neutron-skins of the colliding
nuclei, especially that of the projectile $^{132}Sn$. In the
neutron-rich nuclei, the n/p ratio on the low-density surface is
much higher than that in their interior. Also because of the
$E_{sym}(\rho)\delta^2$ term in the EOS, the isospin-asymmetry in
the low density region is much lower than the supra-normal density
region as long as the symmetry increases with density. In fact, as
shown in Fig. 2 of Ref.~\cite{Li:2005sr}, the isospin-asymmetry of
the low density region can become much higher than the isospin
asymmetry of the reaction system.

It is clearly seen that the dense region can become either
neutron-richer or neutron-poorer with respect to the initial state
depending on the symmetry energy functional $E_{sym}(\rho)$ used. As
long as the symmetry energy increases with the increasing density,
the isospin asymmetry of the supra-normal density region is always
lower than the isospin asymmetry of the reaction system. Thus, even
with radioactive beams, the supra-normal density region can not be
both dense and neutron-rich simultaneously, unlike the situation in
the core of neutron stars, unless the symmetry energy starts
decreasing at high densities. The high density behavior of the
symmetry energy is probably among the most uncertain properties of
dense matter as stressed by~\cite{Kut94,Kut00}. Indeed, some
predictions show that the symmetry energy can decrease with
increasing density above certain density and may even finally
becomes negative. This extreme behavior was first predicted by some
microscopic many-body theories, see e.g.,
Refs.~\cite{Krastev:2006ii,Pan72,Wir88a}. It has also been shown
that the symmetry energy can become negative at various high
densities within the Hartree-Fock approach using the original Gogny
force~\cite{Cha97}, the density-dependent M3Y interaction~
\cite{Kho96,Bas07} and about 2/3 of the 87 Skyrme interactions that
have been widely used in the literature~\cite{Sto03}. The mechanism
leading to and physical meaning of a negative symmetry energy are
the topics of some very recent studies~\cite{Xu09}. It was found
that the spin-isopsin dependence of the three-body force and the
in-medium properties of the short-range tensor force due to the
$\rho$-meson exchange are the main sources of the very uncertain
high-density behavior of the nuclear symmetry energy.

The isospin effects in heavy-ion reactions are determined mainly by
the $E_{sym}(\rho)\delta^2$ term in the EOS. One expects a larger
effect if the isospin-asymmetry is higher. Thus, ideally, one would
like to have situations where both the density and isospin asymmetry
are sufficiently high simultaneously as in the cores of neutron
stars in order to observe the strongest effects due to the symmetry
energy at supra-normal densities. However, since it is the product
of the symmetry energy and the isospin-asymmetry that matters, one
can still probe the symmetry energy at high densities where the
isospin asymmetry is generally low with symmetry energy functionals
that increase with density. Therefore, even if the high density
region may not be as neutron-rich as in neutron stars, heavy-ion
collisions can still be used to probe the symmetry energy at high
densities useful for studying properties of neutron stars.

\begin{figure}[!t]
\centering
\includegraphics[totalheight=3.6in]{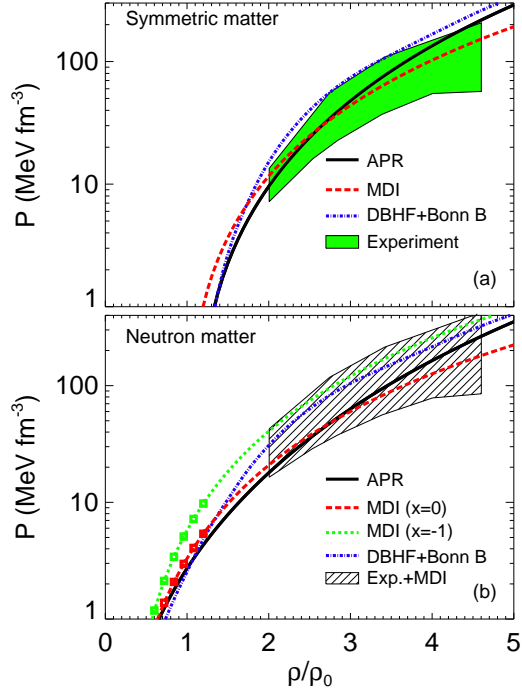}
\vspace{5mm} \caption{(Color online) Pressure as a function of
density for symmetric (upper panel) and pure neutron (lower panel)
matter. The green area in the upper panel is the experimental
constraint on symmetric matter extracted by Danielewicz, Lacey and
Lynch~\cite{Danielewicz:2002pu} from analyzing the collective flow
in relativistic heavy-ion collisions. The corresponding constraint
on the pressure of pure neutron matter, obtained by combining the
flow data and an extrapolation of the symmetry energy functionals
constrained below $1.2\rho_0$ ($\rho_0=0.16 fm^{-3}$) by the isospin
diffusion data, is the shaded black area in the lower panel. Results
are taken partially from Ref.~\cite{Danielewicz:2002pu}.}\label{f3}
\end{figure}

For many astrophysical studies (as those in this chapter), it is
more convenient to express the EOS in terms of the pressure as a
function of density and isospin asymmetry. In Fig.~\ref{f3} we show
pressure as a function of density for two extreme cases: symmetric
(upper panel) and pure neutron matter (lower panel). The green area
in the density range of $\rho_0=[2.0-4.6]$ is the experimental
constraint on the pressure $P_0$ of symmetric nuclear matter
extracted by Danielewicz, Lacey and Lynch from analyzing the
collective flow data from relativistic heavy-ion
collisions~\cite{Danielewicz:2002pu}. The pressure of pure neutron
matter $P_{PNM}=P_0+\rho^2dE_{sym}(\rho)/d\rho$ depends on the
density behavior of the nuclear symmetry energy $E_{sym}(\rho)$.
Since the experimental constraints on the symmetry energy from
terrestrial laboratory experiments are available mostly for
densities less than about $1.2\rho_0$ as indicated by the green and
red squares in the lower panel, which is in contrast to the
constraint on the symmetric EOS that is only available at much
higher densities, the most reliable estimate of the EOS of
neutron-rich matter can thus be obtained by extrapolating the
underlying model EOS for symmetric matter and the symmetry energy in
their respective density ranges to all densities. Shown by the
shaded black area in the lower panel is the resulting best estimate
of the pressure of high density pure neutron matter based on the
predictions from the MDI interaction with $x=0$ and $x=-1$ as the
lower and upper bounds on the symmetry energy and the
flow-constrained symmetric EOS. As one expects and consistent with
the estimate in Ref.~\cite{Danielewicz:2002pu}, the estimated error
bars of the high density pure neutron matter EOS are much wider than
the uncertainty range of the symmetric EOS. For the four
interactions indicated in the figure, their predicted EOSs cannot be
distinguished by the estimated constraint on the high density pure
neutron matter. In addition to the MDI EOS, in Fig.~\ref{f3} we show
results by Akmal et al.~\cite{Akmal:1998cf} with the
$A18+\delta\upsilon+UIX*$ interaction (APR) and recent
Dirac-Brueckner-Hartree-Fock (DBHF)
calculations~\cite{Sammarruca:2008iu} with Bonn B One-Boson-Exchange
(OBE) potential (DBHF+Bonn B)~\cite{Machleidt:1989}. (Older
calculations of the DBHF+Bonn B EOS can be found in
Refs.~\cite{Alonso:2003aq,Krastev:2006ii}.) The saturation
properties of the nuclear equations of state applied in this work
are summarized in Table \ref{tab.2}.

\begin{table}[!t]
\caption{Saturation properties of the nuclear EOSs (for symmetric
nuclear matter) shown in Fig.~\ref{f3}.}
\begin{center}
\begin{tabular}{lccccc}\label{tab.2}
EOS &  $\rho_0$ & $E_s$ & $\kappa$ & $m^*(\rho_0)$ & $E_{sym}(\rho_0)$ \\
    &    $(fm^{-3})$ &  $(MeV)$  &  $(MeV)$   & $(MeV/c^2)$ &
    $(MeV)$ \\
\hline\hline
MDI         & 0.160 & -16.08 & 211.00 & 629.08 &  31.62 \\
APR         & 0.160 & -16.00 & 266.00 & 657.25 &  32.60 \\
DBHF+Bonn B & 0.185 & -16.14 & 259.04 & 610.30 &  33.71 \\
\hline
\end{tabular}
\end{center}
\vspace{3mm}{\small The first column identifies the equation of
state. The remaining columns exhibit the following quantities at the
nuclear saturation density: saturation (baryon) density;
energy-per-particle; compression modulus; nucleon effective mass;
symmetry energy.}
\end{table}

\section{Equations determining the structure of neutron stars}

For completeness, in the following we briefly recall the equations
determining the structure of both static and (rapidly) rotating
neutron stars based on the well-established formalism in the
astrophysical literature. As already mentioned in the introduction,
neutron stars are objects of extremely compressed matter and
therefore proper understanding of their properties requires
application of both General Relativity and the theories of dense
matter. In this respect, neutron stars provide a direct link between
two of the frontiers of modern physics - General Relativity and
strong interactions in dense matter~\cite{Weber:1999a}. The
connection between both branches of physics is provided by
Einstein's field equations
\begin{equation}\label{eq.1}
G^{\mu\nu}=R^{\mu\nu}-\frac{1}{2}g^{\mu\nu}R=8\pi
T^{\mu\nu}(\epsilon,P(\epsilon)),
\end{equation}
$(\mu,\nu=0,1,2,3)$ which couple the Einstein curvature tensor,
$G^{\mu\nu}$, to the energy-momentum tensor,
\begin{equation}\label{eq.2}
T^{\mu\nu} = (\epsilon+P)u^{\mu}u^{\nu}+Pg^{\mu\nu},
\end{equation}
of stellar matter. In the above equations $P$ and $\epsilon$ denote
pressure and mass energy density, while $R^{\mu\nu}$, $g^{\mu\nu}$,
and $R$ denote the Ricci tensor, the metric tensor, and the Ricci
scalar curvature respectively(see e.g.,~\cite{Glendenning:2000a}).
In Eq.~(\ref{eq.2}) $u^\mu$ is the unit time-like four-velocity
satisfying $u^{\mu}u_{\mu}=-1$. The tensor $T^{\mu\nu}$ contains the
EOS of stellar matter in the form $P(\epsilon)$. In general,
Einstein's field equations and those of the nuclear many-body
problem were to be solved simultaneously since the baryons and
quarks follow the geodesics of the curved space-time whose geometry,
determined by the Einstein's field equations, is coupled to the
total mass energy density of matter~\cite{Weber:1999a}. In the case
of neutron stars, as for all astrophysical situations for which the
long-range gravitational force can be separated from the short-range
strong force, the deviation from flat space-time at the length-scale
of the strong interactions ($\sim 1fm$) is practically zero up to
the highest densities achieved in the neutron star interiors. (This
is not to be confused with the global length-scale of neutron stars
($\sim 10km$) for which $M/R\sim 0.3$ depending on the star's mass
(in units $c=G=1$ so that $M_{\odot}\approx 1.475km$).) In other
words, gravity curves space-time only on a macroscopic scale but to
a very good approximation leaves it flat on a microscopic scale. To
achieve an appreciable curvature on a microscopic level at which the
strong interactions dominate the particle dynamics mass densities
greater than $\sim 10^{40} g\hspace{1mm}cm^{-3}$ would be
necessary~\cite{Weber:1999a,Thorne1966a}. Under this circumstances
the problem of constructing models of neutron stars separates into
two distinct tasks. First, the short-range effects of the nuclear
forces are described by the principles of many-body nuclear physics
in a local inertial frame (co-moving proper reference frame) in
which space-time is flat. Second, the coupling between the
long-range gravitational force and matter is accounted for by
solving the general relativistic equations for the gravitational
field described by the curvature of space-time, leading to the
global structure of stellar configurations.

\subsection{Static stars}

In the case of spherically symmetric static (non-rotating) stars the
metric has the famous Schwarzschild form:
\begin{equation}\label{eq.3}
ds^2=-e^{2\phi(r)}dt^2+e^{2\Lambda(r)}dr^2+r^2(d\theta^2+\sin^2\theta
d\phi^2),
\end{equation}
$(c=G=1)$ where the metric functions $\phi(r)$ and $\Lambda(r)$ are
given by:
\begin{equation}\label{eq.4}
e^{2\Lambda(r)}=(1-\gamma(r))^{-1},
\end{equation}
\begin{equation}\label{eq.5}
e^{2\phi(r)}=e^{-2\Lambda(r)}=(1-\gamma(r))\quad r>R_{star},
\end{equation}
with
\begin{equation}\label{eq.6}
\gamma(r)=\left\{
\begin{array}{l l}
\frac{2m(r)}{r} & \quad \mbox{if $r<R_{star}$}\\\\
\frac{2M}{r} & \quad \mbox{if $r>R_{star}$}
\end{array}
\right.
\end{equation}
For a static star Einstein's field equations (Eq.~(\ref{eq.1}))
reduce then to the familiar Tolman-Oppenheimer-Volkoff equation
(TOV)~\cite{Tolman:1939jz,PhysRev.55.374}:
\begin{equation}\label{eq.7}
\frac{dP(r)}{dr}=-\frac{\epsilon(r)m(r)}{r^2}
\left[1+\frac{P(r)}{\epsilon(r)}\right]
\left[1+\frac{4\pi{r^3}p(r)}{m(r)}\right]
\left[1-\frac{2m(r)}{r}\right]^{-1}
\end{equation}
where the gravitational mass within a sphere of radius $r$ is
determined by
\begin{equation}\label{eq.8}
\frac{dm(r)}{dr}=4\pi\epsilon(r)r^{2}
\end{equation}
The metric function $\phi(r)$ is determined through the following
differential equation:
\begin{equation}\label{eq.9}
\frac{d\phi(r)}{dr}=-\frac{1}{\epsilon(r)+P(r)}\frac{dP(r)}{dr},
\end{equation}
with the boundary condition at $r=R$
\begin{equation}\label{eq.10}
\phi(r=R)=\frac{1}{2}\ln\left(1-\frac{2M}{R}\right)
\end{equation}

To proceed to the solution of these equations, it is necessary to
provide the EOS of stellar matter in the form $P(\epsilon)$.
Starting from some central energy density $\epsilon_c=\epsilon(0)$
at the center of the star $(r=0)$, and with the initial condition
$m(0)=0$, the above equations can be integrated outward until the
pressure vanishes, signifying that the stellar edge is reached. Some
care should be taken at $r=0$ since, as seen above, the TOV equation
is singular there. The point $r=R$  where the pressure vanishes
defines the radius of the star and
$M=m(R)=4\pi\int_0^R\epsilon(r')r'^2dr'$ its gravitational mass.

For a given EOS, there is a unique relationship between the stellar
mass and the central density $\epsilon_c$. Thus, for a particular
EOS, there is a unique sequence of stars parameterized by the
central density (or equivalently the central pressure $P(0)$).

\subsection{Rotating stars}

Equations of stellar structure of (rapidly) rotating neutron stars
are considerably more complex than those of spherically symmetric
stars~\cite{Weber:1999a}. These complications arise due to the
rotational deformations in rotating stars (i.e., flattening at the
poles and bulging at the equator), which lead to a dependence of the
star's metric on the polar coordinate $\theta$. In addition,
rotation stabilizes the star against gravitational collapse and
therefore rotating neuron stars are more massive than static ones. A
larger mass, however, causes greater curvature of space-time. This
renders the metric functions frequency-dependent. Finally, the
general relativistic effect of dragging the local inertial frames
implies the occurrence of an additional non-diagonal term,
$g^{t\phi}$, in the metric tensor $g^{\mu\nu}$. This term imposes a
self-consistency condition on the stellar structure equations, since
the degree at which the local inertial frames are dragged along by
the star, is determined by the initially unknown stellar properties
like mass and rotational frequency~\cite{Weber:1999a}.

The structure equations of rapidly rotating neutron stars have been
computed with the $RNS$\footnote{Thanks to Nikolaos Stergioulas the
$RNS$ code is available as a public domain program at
http://www.gravity.phys.uwm.edu/rns/} code developed and made
available to the public by Nikolaos
Stergioulas~\cite{Stergioulas:1994ea}. Here we outline briefly the
equations solved by the $RNS$ code. The coordinates of the
stationary, axial symmetric space-time used to model a (rapidly)
rotating neutron star are defined through a generalization of
Bardeen's metric~\cite{Stergioulas:2003yp}:
\begin{eqnarray}\label{eq.11}
ds^2&=&-e^{\gamma+\rho}dt^2+e^{2\alpha}(dr^2+r^2d\theta^2)\nonumber\\
&+&e^{\gamma-\rho}r^2\sin^2\theta(d\phi-\omega dt^2),
\end{eqnarray}
where the metric potentials $\gamma$, $\rho$, $\alpha$, and the
angular velocity of the stellar fluid relative to the local inertial
frame, $\omega$, are functions of the quasi-isotropic radial
coordinates, $r$, and the polar angle $\theta$ only. The matter
inside a rigidly rotating star is approximated as a perfect
fluid~\cite{Stergioulas:2003yp}, whose energy momentum tensor is
given by Eq.~(\ref{eq.2}). The proper velocity of matter,
$\upsilon$, relative to the local Zero Angular Momentum Observer
(ZAMO)~\cite{2002A&A...382..939O} is defined as
\begin{equation}\label{eq.12}
\upsilon=r\sin(\theta)(\Omega-\omega)e^{-\rho(r)}
\end{equation}
with $\Omega=u^3/u^0$ the angular velocity of a fluid element. The
four-velocity is given by
\begin{equation}\label{eq.13}
u^{\mu}=\frac{e^{-(\gamma+\rho)/2}}{\sqrt{(1-\upsilon^2)}}(1,0,0,\Omega)
\end{equation}
In the above equation the function $(\gamma+\rho)/2$ represents the
relativistic generalization of the Newtonian gravitational
potential, while $\exp[(\gamma+\rho)/2]$ is a time dilation factor
between an observer moving with angular velocity $\omega$ and one at
infinity. Substitution of Eq.~(\ref{eq.13}) into Einstein's fields
equations projected onto the ZAMO reference frame gives three
elliptic partial differential equations for the metric potentials
$\gamma$, $\rho$, and $\omega$, and two linear ordinary differential
equations for the metric potential $\alpha$. Technically, the
elliptic differential equations for the metric functions are
converted into integral equations which are then solved iteratively
applying Green's function
approach~\cite{1989MNRAS.237..355K,Stergioulas:2003yp}.

From the relativistic equations of motion, the equations of
hydrostatic equilibrium for a barotropic fluid may be obtained
as~\cite{Stergioulas:2003yp,2002A&A...382..939O}:
\begin{equation}\label{eq.14}
h(P)-h_p=\frac{1}{2}[\omega_p+\rho_p-\gamma-\rho-\ln(1-\upsilon^2)+A^2(\omega-\Omega_c)^2],
\end{equation}
with $h(P)$ the specific enthalpy, $P_p$ the re-scaled pressure,
$h_p$ the specific enthalpy at the pole, $\gamma_p$ and $\rho_p$ the
values of the metric potentials at the pole, $\Omega_c=r_e\Omega$,
and $A$ a rotational constant~\cite{Stergioulas:2003yp}. The
subscripts $p$, $e$, and $c$ label the corresponding quantities at
the pole, equator and center respectively. The $RNS$ code solves
iteratively the integral equations for $\rho$, $\gamma$ and
$\omega$, and the ordinary differential equation for the metric
function $\alpha$ coupled with Eq.~(\ref{eq.14}) and the equations
for hydrostatic equilibrium at the stellar center and equator (given
$h(P_c)$ and $h(P_e)=0$) to obtain $\rho$, $\gamma$, $\alpha$,
$\omega$, the equatorial coordinate radius $r_e$, angular velocity
$\Omega$, energy density $\epsilon$, and pressure $P$ throughout the
star.

\section{Constraining global properties of pulsars}

We construct one-parameter 2-D stationary configurations of
(rapidly) rotating neutron stars employing several nucleonic EOSs
and the $RNS$ code. We assume a simple model of stellar matter of
nucleons and light leptons (electrons and muons) in
beta-equilibrium. Below the baryon density of approximately
$0.07fm^{-3}$ the equations of state applied here are supplemented
by a crustal EOS, which is more suitable for the low density
regime. Namely, we apply the EOS by Pethick et al.~\cite{PRL1995}
for the inner crust and the one by Haensel and
Pichon~\cite{HP1994} for the outer crust. In the core we assume a
continuous functional for the EOSs employed in this work. (See
Ref.~\cite{Krastev:2006ii} for a detailed description of the
extrapolation procedure for the DBHF+Bonn B EOS.)

\subsection{Keplerian (and static) sequences}

\begin{figure}[!t]
\centering
\includegraphics[totalheight=3.6in]{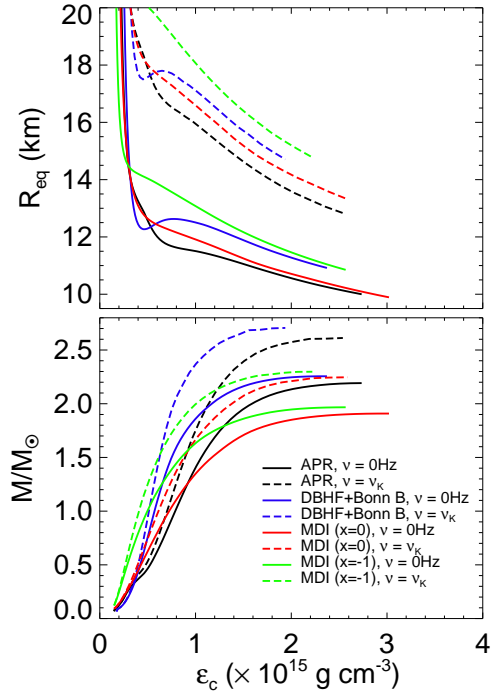}
\vspace{5mm} \caption{(Color online) Neutron star masses and radii.
Neutron star equatorial radii (upper panel) and total gravitational
mass (lower panel) versus central energy density $\epsilon_c$. Both
static (solid lines) and Keplerian (broken lines) models are shown.
Taken from Ref. \cite{KLW2}.}\label{f4}
\end{figure}

In what follows we examine the effect of ultra-fast rotation at the
Kepler frequency on the neutron star gravitational masses and radii.
The equilibrium configurations for both static and (rapidly)
rotating neutron stars are parameterized in terms of the central
mass energy density, $\epsilon_c=\epsilon(0)$ (or equivalently
central pressure, $P_c=P(0)$). This functional dependence is shown
in Fig.~\ref{f4}, where we display the stellar {\it equatorial}
radius (upper frame) and total gravitational mass (lower frame)
versus central energy density for the EOSs applied in this work.
Predictions for both static and maximally rotating models are shown.
\begin{figure}[!t]
\centering
\includegraphics[totalheight=2.6in]{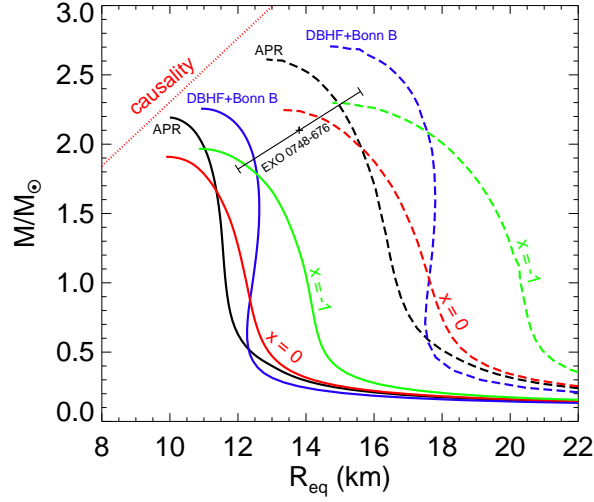}
\caption{(Color online) Mass-radius relation. Both static (solid
lines) and Keplerian (broken lines) sequences are shown. The
$1-\sigma$ error bar corresponds to the measurement of the mass and
radius of EXO 0748-676~\cite{Ozel:2006km}. Taken from Ref.
\cite{KLW2}.}\label{f5}
\end{figure}
\begin{figure}[!t]
\centering
\includegraphics[totalheight=2.4in]{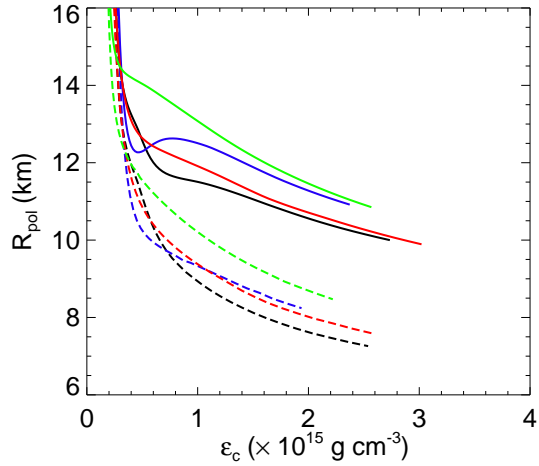}
\caption{(Color online) Neutron-star polar radius versus central
energy density. Both static (solid lines) and Keplerian (broken
lines) sequences are shown. The labeling of the curves is the same
as in Fig.~\ref{f4}. Taken from Ref. \cite{KLW2}.}\label{f6}
\end{figure}
The main feature of the mass-density plot is that there exists a
maximum value of the gravitational mass of a neutron star that a
given EOS can support~\cite{Weber:1999a}. This holds for both static
and (rapidly) rotating stars. The sequences shown in Fig.~\ref{f4}
terminate at the ''maximum mass" point. Comparing the results for
static and rotating stars, it is seen clearly that the rapid
rotation increases noticeably the mass that can be supported against
collapse while lowering the central density of the maximum-mass
configuration. This is what one should expect, since, as already
mentioned, rotation stabilizes the star against the gravitational
pull providing an extra (centrifugal) repulsion. The rotational
effect on the mass-radius relation is illustrated in Fig.~\ref{f5}
where the gravitational mass is given as a function of the
circumferential radius. For rapid rotation at the Kepler frequency,
a mass increase up to $\sim 17\%$ (Table~\ref{tab.4}) is obtained,
depending on the EOS. The equatorial radius increases by several
kilometers, while the polar radius decreases by several kilometers
(see Fig.~\ref{f6}) leading to an overall oblate shape of the
rotating star. Table \ref{tab.3} summarizes the properties (masses,
radii and central energy densities) of the maximum-mass non-rotating
neutron star configurations. Our studies on the effect of rapid
rotation on the upper mass limits for the four EOSs considered in
this chapter are presented in Table \ref{tab.4}. In each case the
upper mass limit is obtained for a model at the mass-shedding limit
where $\nu=\nu_k$, with central density $\sim 15\%$ below that of
the static model with the largest mass. These findings are
consistent with those in Refs. \cite{1984Natur.312..255F} and
\cite{Stergioulas:1994ea}. Table \ref{tab.4} also provides an
estimate of the upper limiting rotation rate of a neutron star. In
general, softer EOSs permit larger rotational frequencies since the
resulting stellar models are more centrally condensed, see e.g.,
Ref. \cite{1984Natur.312..255F}. In the last column of Table
\ref{tab.4} we show the Kepler frequencies computed via the
empirical relation
\begin{table}[!t]
\caption{Maximum-mass static (non-rotating) models.}
\begin{center}
\begin{tabular}{lccc}\label{tab.3}
EOS &  $M_{max}(M_{\odot})$ & $R(km)$ & $\epsilon_c(\times 10^{15}g\hspace{1mm}cm^{-3})$\\
\hline\hline
MDI(x=0)       & 1.91    &  9.89  & 3.02\\
APR            & 2.19    &  9.98  & 2.73\\
MDI(x=-1)      & 1.97    & 10.85  & 2.57\\
DBHF+Bonn B    & 2.26    & 10.91  & 2.37\\
\hline
\end{tabular}
\end{center}
{\small The first column identifies the equation of state. The
remaining columns exhibit the following quantities for the static
models with maximum gravitational mass: gravitational mass; radius;
central mass energy density.}
\end{table}
\begin{table}[!t]
\caption{Maximum-mass rapidly rotating models at the Kepler
frequency $\nu=\nu_k$.}
\begin{center}
\begin{tabular}{lccccc}\label{tab.4}
EOS &  $M_{max}(M_{\odot})$ & Increase (\%) & $\epsilon_c(\times
10^{15}g\hspace{1mm}cm^{-3})$ & $\nu_k(Hz)$
& $\nu_k^{FIP}(Hz)$\\
\hline\hline
MDI(x=0)       & 2.25    &  15  & 2.59 & 1742 & 1610\\
APR            & 2.61    &  17  & 2.53 & 1963 & 1699\\
MDI(x=-1)      & 2.30    &  14  & 2.21 & 1512 & 1423\\
DBHF+Bonn B    & 2.71    &  17  & 1.94 & 1644 & 1510\\
\hline
\end{tabular}
\end{center}
{\small The first column identifies the equation of state. The
remaining columns exhibit the following quantities for the maximally
rotating models with maximum gravitational mass: gravitational mass;
its percentage increase over the maximum gravitational mass of
static models; central mass energy density; maximum rotational
frequency; Kepler frequency as computed via the empirical relation
given by Eq. (\ref{eq.15})~\cite{Friedman:1989}.}
\end{table}
\begin{equation}\label{eq.15}
\frac{\Omega_k}{10^4s^{-1}}=0.72\left(\frac{M_s}{M_{\odot}}\right)^{1/2}\left(\frac{R_s}{10km}\right)^{-3/2}
\end{equation}
proposed in Ref. \cite{Friedman:1989}. The uncertainty of
Eq.~(\ref{eq.15}) is $\sim 10\%$ (see Ref.~\cite{HZ1989} for an
improved version of the empirical formula). At the time of
constructing the above relation Friedman et al.~\cite{Friedman:1989}
did not consider a then unknown class of minimum period EOSs
\cite{Stergioulas:1996} which explains why the numbers in the last
column of Table \ref{tab.4} exhibit larger deviation from the exact
numerical solutions (in column five). This is particularly
pronounced for the APR EOS for which the approximated Kepler
frequency deviates $\sim 14\%$ from the exact solution. (Note that
the APR EOS has the lowest period among the EOSs considered here.)

\begin{figure}[!t]
\centering
\includegraphics[totalheight=2.6in]{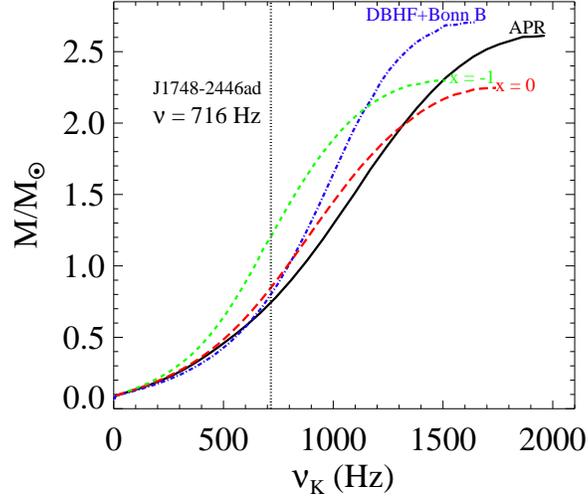}
\caption{(Color online) Mass versus Keplerian (mass-shedding)
frequency $\nu_k$. Adapted from Ref. \cite{KLW2}.}\label{f7}
\end{figure}

Fig.~\ref{f7} displays the neutron star gravitational mass as a
function of the Kepler frequency, $\nu_k$. The vertical doted line
corresponds to the frequency of the fastest neutron star presently
known (J1748-2446ad \cite{Hessels:2006ze}). The figure suggests that
the higher the rotational frequency, the larger the mass for the
possible stable neutron star configurations. We discuss this further
in the next subsection.

\subsection{Rotation at various frequencies}

\begin{figure}[!t]
\centering
\includegraphics[totalheight=2.4in]{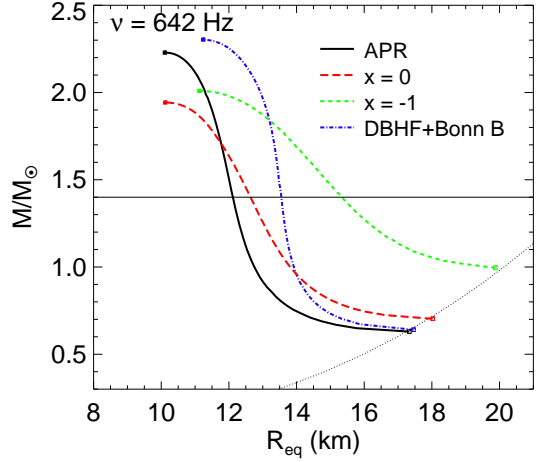}
\caption{(Color online) Gravitational mass versus circumferential
radius for neutron stars rotating at $\nu=642Hz$. See text for
details.}\label{f8}
\end{figure}

\begin{figure}[!t]
\centering
\includegraphics[totalheight=2.4in]{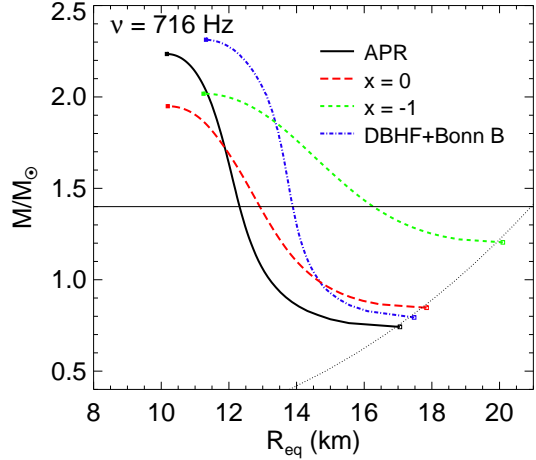}
\caption{(Color online) Gravitational mass versus circumferential
radius for neutron stars rotating at $\nu=716Hz$.}\label{f9}
\end{figure}

\begin{figure}[!t]
\centering
\includegraphics[totalheight=2.4in]{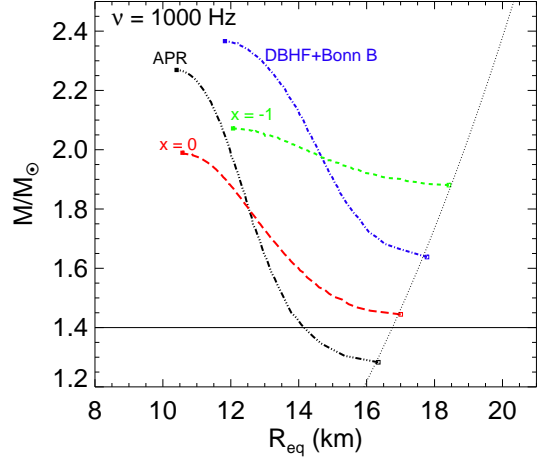}
\caption{(Color online) Gravitational mass versus circumferential
radius for neutron star models  rotating at
$\nu=1000Hz$.}\label{f10}
\end{figure}

In this subsection we study neutron stars rotating at various
(fixed) frequencies. Stability with respect to the mass-shedding
from equator implies that at a given gravitational mass the
equatorial radius $R_{eq}$ should be smaller than $R_{eq}^{max}$
corresponding to the Keplerian limit~\cite{Bejger:2006hn}. The value
of $R_{eq}^{max}$ results from the condition that the frequency of a
test particle at circular equatorial orbit of radius $R_{eq}^{max}$
just above the equator of the actual rotating star is equal to the
rotational frequency of the star. As reported by Bejger et
al.~\cite{Bejger:2006hn} the relation between $M$ and $R_{eq}$ at
the ``mass-shedding point'' is very well approximated by the
expression for the orbital frequency for a test particle orbiting at
$r=R_{eq}$ in the Schwarzschild space-time created by a spherical
mass. The formula satisfying $\nu_{orb}^{Schw.}=\nu$, represented by
the dotted line in Figs.~\ref{f8}, \ref{f9} and \ref{f10} is given
by
\begin{equation}\label{eq.16}
\frac{1}{2\pi}\left(\frac{GM}{R_{eq}^3}\right)=\nu,
\end{equation}
where $\nu=642Hz$ in Fig.~\ref{f8}, $\nu=716Hz$ in Fig.~\ref{f9},
and $\nu=1000Hz$ in Fig.~\ref{f10} respectively. (Rotational
frequencies $642Hz$ and $716Hz$ are the spinning rates of the two
fastest pulsars PSR1937+214 \cite{Backer:1982} and J1748-2446ad
\cite{Hessels:2006ze}.) This formula for the Schwarzschild metric
coincides with the one obtained in Newtonian gravity for a point
mass $M$ \cite{Bejger:2006hn}. Eq.~(\ref{eq.16}) implies
\begin{equation}\label{eq.17}
R_{max}=\chi\left(\frac{M}{1.4M_{\odot}}\right)^{1/3}km,
\end{equation}
with $\chi=22.52$ for rotational frequency $\nu=642Hz$
(Fig.~\ref{f8}), $\chi=20.94$ for $\nu=716Hz$ (Fig.~\ref{f9}), and
$\chi=16.76$ for $\nu=1000Hz$ (Fig.~\ref{f10}).

In Figs.~\ref{f8}-\ref{f10} we observe that the range of the
allowed masses supported by a given EOS for rapidly rotating
neutron stars becomes narrower than the one of static
configurations. This effect becomes stronger with increasing
frequency and depends upon the EOS. For instance, for models
rotating at $1000Hz$ (Fig.~\ref{f10}) for the $x=-1$ EOS the
allowed mass range is $\sim 0.2M_{\odot}$. As already discussed in
the literature \cite{Bejger:2006hn}, this observation could
explain why such rapidly rotating neutron stars are so rare --
their allowed masses fall within a very narrow range.

\section{Neutron star moment of inertia}

In this section we turn our attention to the moment of inertia of
neutron stars. Such studies are important and timely as they are
related to astrophysical observations in the near future. In
particular, the moment of inertia of pulsar A in the extremely
relativistic neutron star binary PSR J0737-3039 \cite{Burgay2003}
may be determined in a few years through detailed measurements of
the periastron advance \cite{BBH2005}.

Employing the EOSs described briefly in section 2, we compute the
neutron star moment of inertia with the $RNS$ code. It solves the
hydrostatic and Einstein's field equations for mass distributions
rotating rigidly under the assumption of stationary and axial
symmetry about the rotational axis, and reflectional symmetry about
the equatorial plane. $RNS$ calculates the angular momentum $J$ as
\cite{Stergioulas:2003yp}
\begin{equation}\label{eq.24}
J=\int T^{\mu\nu}\xi^{\nu}_{(\phi)}dV,
\end{equation}
where $T^{\mu\nu}$ is the energy-momentum tensor of stellar matter
\begin{equation}\label{eq.25}
T^{\mu\nu} = (\epsilon+P)u^{\mu}u^{\nu}+Pg^{\mu\nu},
\end{equation}
$\xi^{\nu}_{(\phi)}$ is the Killing vector in azimuthal direction
reflecting axial symmetry, and $dV=\sqrt{-g}d^3x$ is a proper
3-volume element ($g\equiv \det(g_{\alpha\beta})$ is the determinant
of the 3-metric). In Eq.~(\ref{eq.25}) $P$ is the pressure,
$\epsilon$ is the mass-energy density, and $u^\mu$ is the unit
time-like four-velocity satisfying $u^{\mu}u_{\mu}=-1$. For
axial-symmetric stars it takes the form $u^{\mu}=u^t(1,0,0,\Omega)$,
where $\Omega$ is the star's angular velocity. Under this condition
Eq.~(\ref{eq.24}) reduces to
\begin{equation}\label{eq.26}
J=\int (\epsilon+P)u^t(g_{\phi\phi}u^{\phi}+g_{\phi
t}u^t)\sqrt{-g}d^3x
\end{equation}
It should be noted that the moment of inertia cannot be calculated
directly as an integral quantity over the source of gravitational
field \cite{Stergioulas:2003yp}. In addition, there exists no unique
generalization of the Newtonian definition of the moment of inertia
in General Relativity and therefore $I=J/\Omega$ is a natural choice
for calculating this important quantity.

For rotational frequencies much lower than the Kepler frequency (the
highest possible rotational rate supported by a given EOS), i.e.
$\nu/\nu_k<<1$ ($\nu=\Omega/(2\pi)$), the deviations from spherical
symmetry are very small, so that the moment of inertia can be
approximated from spherical stellar models. In what follows we
review briefly this slow-rotation approximation, see e.g. Ref.
\cite{Hartle:1967he}. In the slow-rotational limit the metric can be
written in spherical coordinates as (in geometrized units $G=c=1$)
\begin{equation}\label{eq.27}
ds^2=-e^{2\phi(r)}dt^2+\left(1-\frac{2m(r)}{r}\right)^{-1}dr^2-2\omega
r^2\sin^2\theta dt d\phi+r^2(d\theta^2+\sin^2\theta d\phi^2)
\end{equation}
In the above equation $m(r)$ is the total gravitational mass within
radius $r$ satisfying the usual equation
\begin{equation}\label{eq.28}
\frac{dm(r)}{dr}=4\pi\epsilon(r)r^{2}
\end{equation}
and $\omega(r)\equiv(d\phi/dt)_{ZAMO}$ is the Lense-Thirring angular
velocity of a zero-angular-momentum observer (ZAMO). Up to first
order in $\omega$ all metric functions remain spherically symmetric
and depend only on $r$ \cite{Morrison:2004}. In the stellar interior
the Einstein's field equations reduce to
\begin{equation}\label{eq.29}
\frac{d\phi(r)}{dr}=m(r)\left[1+\frac{4\pi
r^3P(r)}{m(r)}\right]\left[1-\frac{2m(r)}{r}\right]^{-1}\quad
(r<R_{star})
\end{equation}
and
\begin{equation}\label{eq.30}
\frac{1}{r^3}\frac{d}{dr}\left(r^4j(r)\frac{d\bar{\omega}(r)}{dr}\right)+4\frac{dj(r)}{dr}\bar{\omega}(r)=0\quad
(r<R_{star}),
\end{equation}
with $\bar{\omega}\equiv \Omega-\omega$ the dragging angular
velocity (the angular velocity of the star relative to a local
inertial frame rotating at $\omega$) and
\begin{equation}\label{eq.31}
j\equiv\left(1-\frac{2m(r)}{r}\right)^{1/2}e^{-\phi(r)}
\end{equation}
Outside the star the metric functions become
\begin{equation}\label{eq.32}
e^{2\phi}=\left(1-\frac{2M}{r}\right)\quad (r>R_{star})
\end{equation}
and
\begin{equation}\label{eq.33}
\omega=\frac{2J}{r^3}\quad (r>R_{star}),
\end{equation}
where $M=m(r=R)=4\pi\int_0^R\epsilon(r')r'^2dr'$ is the total
gravitational mass and $R$ is the stellar radius defined as the
radius at which the pressure drops to zero ($P(r=R)=0$). At the
star's surface the interior and exterior solutions are matched by
satisfying the appropriate boundary conditions
\begin{equation}\label{eq.34}
\bar{\omega}(R)=\Omega-\frac{R}{3}\left(\frac{d\bar{\omega}}{dr}\right)_{r=R}
\end{equation}
and
\begin{equation}\label{eq.35}
\phi(r)=\frac{1}{2}\ln\left(1-\frac{2M}{R}\right)
\end{equation}
The moment of inertia $I=J/\Omega$ then can be computed from
Eq.~(\ref{eq.26}). With $\Omega=u^{\phi}/u^{t}$ and retaining only
first order terms in $\omega$ and $\Omega$, the moment of inertia
reads \cite{Morrison:2004,Lattimer:2000kb}
\begin{equation}\label{eq.36}
I\approx\frac{8\pi}{3}\int^R_0(\epsilon+P)e^{-\phi(r)}\left[1-\frac{2m(r)}{r}\right]^{-1}
\frac{\bar{\omega}}{\Omega}r^4dr
\end{equation}
This slow-rotation approximation for the neutron-star moment of
inertia neglects deviations from spherical symmetry and is
independent of the angular velocity $\Omega$ \cite{Morrison:2004}.
For neutron stars with masses greater than $1M_{\odot}$ Lattimer and
Schutz~\cite{Lattimer:2005} found that, for slow-rotations, the
moments of inertia computed through the above formalism
(Eq.~(\ref{eq.36})) can be approximated very well by the following
empirical relation:
\begin{equation}\label{eq.37}
I\approx(0.237\pm
0.008)MR^2\left[1+4.2\frac{Mkm}{M_{\odot}R}+90\left(\frac{Mkm}{M_{\odot}R}\right)^4\right]
\end{equation}
The above equation is shown \cite{Lattimer:2005} to hold for a wide
class of EOSs except for ones with appreciable degree of softening,
usually indicated by achieving a maximum mass of $\sim 1.6M_{\odot}$
or less. Since none of the EOSs employed in this paper exhibit such
pronounced softening, Eq.~(\ref{eq.37}) is a good approximation for
the momenta of inertia of {\it slowly} rotating stars.

\subsection{Slow rotation}

\begin{figure}[t!]
\centering
\includegraphics[totalheight=2.4in]{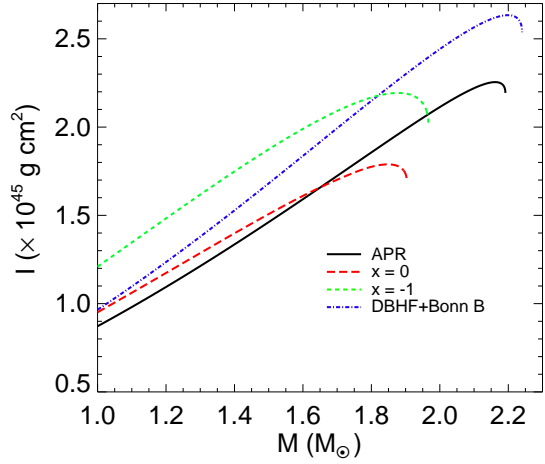}
\vspace{5mm} \caption{(Color online) Total moment of inertia of
neutron stars estimated with Eq.~(\ref{eq.37}). Taken from
Ref.~\cite{WKL:2008ApJ}.} \label{f11}
\end{figure}
\begin{figure}[!t]
\centering
\includegraphics[totalheight=2.4in]{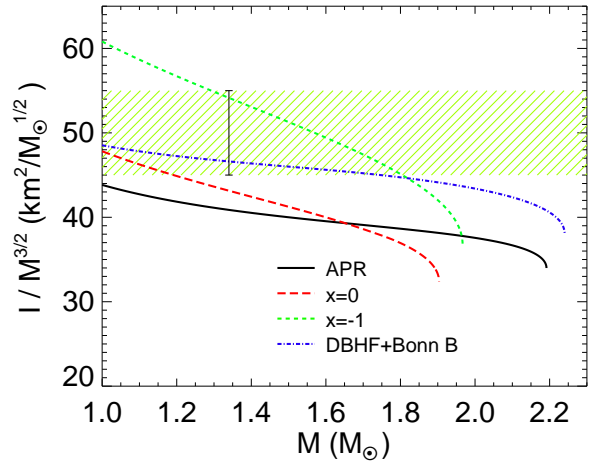}
\vspace{5mm} \caption{(Color online) The moment of inertia scaled by
$M^{3/2}$ as a function of the stellar mass $M$. The shaded band
illustrates a 10\% error of hypothetical $I/M^{3/2}$ measurement of
50 $km^2$ $M_{\odot}^{-1/2}$. The error bar shows the specific case
in which the mass is $1.34M_{\odot}$ (after Lattimer and Schutz
\cite{Lattimer:2005}). Taken from Ref.~\cite{WKL:2008ApJ}.}
\label{f12}
\end{figure}

If the rotational frequency is much smaller than the Kepler
frequency, the deviations from spherical symmetry are negligible and
the moment of inertia can be calculated applying the slow-rotation
approximation discussed briefly above. For this case Lattimer and
Schutz \cite{Lattimer:2005} showed that the moment of inertia can be
very well approximated by Eq.~(\ref{eq.37}). In Fig.~\ref{f11} we
display the moment of inertia as a function of stellar mass for
slowly rotating neutron stars as computed with the empirical
relation Eq. (\ref{eq.37}). As shown in Fig.~\ref{f5}, above $\sim
1.0M_{\odot}$ the neutron star radius remains approximately constant
before reaching the maximum mass supported by a given EOS. The
moment of inertia ($I\sim MR^2$) thus increases almost linearly with
stellar mass for all models. Right before the maximum mass is
achieved, the neutron star radius starts to decrease
(Fig.~\ref{f5}), which causes the sharp drop in the moment of
inertia observed in Fig.~\ref{f11}. Since $I$ is proportional to the
mass and the square of the radius, it is more sensitive to the
density dependence of the nuclear symmetry energy, which determines
the neutron star radius. Here we recall that the $x=-1$ EOS has much
stiffer symmetry energy (with respect to the one of the $x=0$ EOS),
which results in neutron star models with larger radii and, in turn,
momenta of inertia. For instance, for a ``canonical'' neutron star
($M=1.4M_{\odot}$), the difference in the moment of inertia is more
than $30\%$ with the $x=0$ and the $x=-1$ EOSs. In Fig.~\ref{f12} we
take another view of the moment of inertia where $I$ is scaled by
$M^{3/2}$ as a function of the stellar mass (after
\cite{Lattimer:2005}).

The discovery of the extremely relativistic binary pulsar PSR
J0737-3039A,B provides an unprecedented opportunity to test General
Relativity and physics of pulsars \cite{Burgay2003}. Lattimer and
Schutz~\cite{Lattimer:2005} estimated that the moment of inertia of
the A component of the system should be measurable with an accuracy
of about 10\%. Given that the masses of both stars are already
accurately determined by observations, a measurement of the moment
of inertia of even one neutron star could have enormous importance
for the neutron star physics \cite{Lattimer:2005}. (The significance
of such a measurement is illustrated in Fig.~\ref{f12}. As pointed
by \cite{Lattimer:2005}, it is clear that very few EOSs would
survive these constraints.) Thus, theoretical predictions of the
moment of inertia are very timely. Calculations of the moment of
inertia of pulsar A ($M_A=1.338M_{\odot}$, $\nu_A=44.05Hz$) have
been reported by Morrison et al.~\cite{Morrison:2004} and Bejger et
al.~\cite{BBH2005}.
\begin{table}[t!]
\caption{Numerical results for PSR J0737-3039A
($M_A=1.338M_{\odot}$, $\nu_A=44.05Hz$).}
\begin{center}
\begin{tabular}{lcccc}\label{tab.5}
EOS &  $\epsilon_c(\times 10^{14}g$ $cm^{-3})$ & $R_{eq}(km)$ & $I(\times 10^{45}g$ $cm^2)$ & $I^{LS}(\times 10^{45}g$ $cm^2)$\\
\hline\hline
MDI(x=-1)    & 7.04 & 13.75 (13.64) & 1.63 & 1.67\\
DBHF+Bonn B  & 7.34 & 12.56 (12.47) & 1.57 & 1.43\\
MDI(x=0)     & 9.85 & 12.00 (11.90) & 1.30 & 1.34\\
APR          & 9.58 & 11.60 (11.52) & 1.25 & 1.26\\
\hline
\end{tabular}
\end{center}
{\small The first column identifies the equation of state. The
remaining columns exhibit the following quantities: central
mass-energy density, equatorial radius (the numbers in the
parenthesis are the radii of the spherical models; the deviations
from sphericity due to rotation are $\sim 1\%$), total moment of
inertia, total moment of inertia $I^{LS}$ as computed with
Eq.~(\ref{eq.37}).}
\end{table}
In Table~\ref{tab.5} we show the moment of inertia (and other
selected quantities) of PSR J0737-3039A computed with the $RNS$ code
using the EOSs employed in this study. Our results with the APR EOS
are in very good agreement with those by \cite{Morrison:2004}
($I^{APR}=1.24\times 10^{45}g$ $cm^2$) and \cite{BBH2005}
($I^{APR}=1.23\times 10^{45}g$ $cm^2$). In the last column of
Table~\ref{tab.5} we also include results computed with the
empirical relation (Eq.~(\ref{eq.37})). From a comparison with the
results from the exact numerical calculation we conclude that
Eq.~(\ref{eq.37}) is an excellent approximation for the moment of
inertia of slowly-rotating neutron stars. (The average uncertainty
of Eq.~(\ref{eq.37}) is $\sim 2\%$, except for the DBHF+BonnB EOS
for which it is $\sim 8\%$.) Our results (with the MDI EOS) allowed
us to constrain the moment of inertia of pulsar A to be in the range
$I=(1.30-1.63)\times 10^{45}(g$ $cm^2)$.

\subsection{Rapid rotation}

\begin{figure}[!t]
\centering
\includegraphics[totalheight=2.3in]{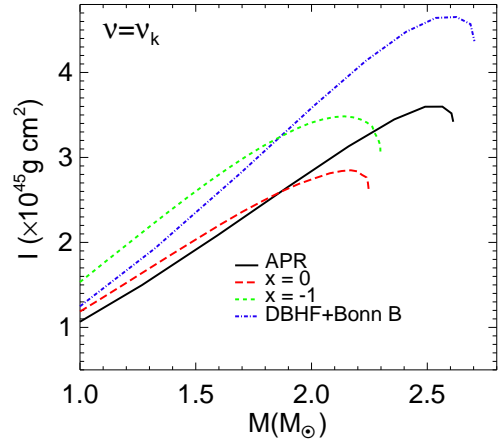}
\vspace{5mm} \caption{(Color online) Total moment of inertia for
Keplerian models. The neutron star sequences are computed with the
$RNS$ code. Taken from Ref. \cite{WKL:2008ApJ}.} \label{f13}
\end{figure}

\begin{figure}[!t]
\centering
\includegraphics[totalheight=4.2in]{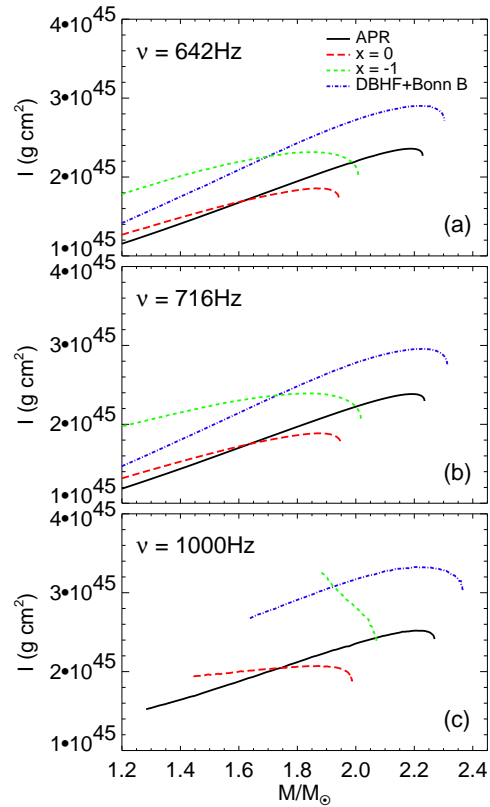}
\vspace{5mm} \caption{(Color online) Total moment of inertia as a
function of stellar mass for models rotating at 642Hz (a), 716Hz
(b), and 1000Hz (c). Data is partially taken from
Ref.~\cite{WKL:2008ApJ}.} \label{f14}
\end{figure}

\begin{figure}[!t]
\centering
\includegraphics[totalheight=2.3in]{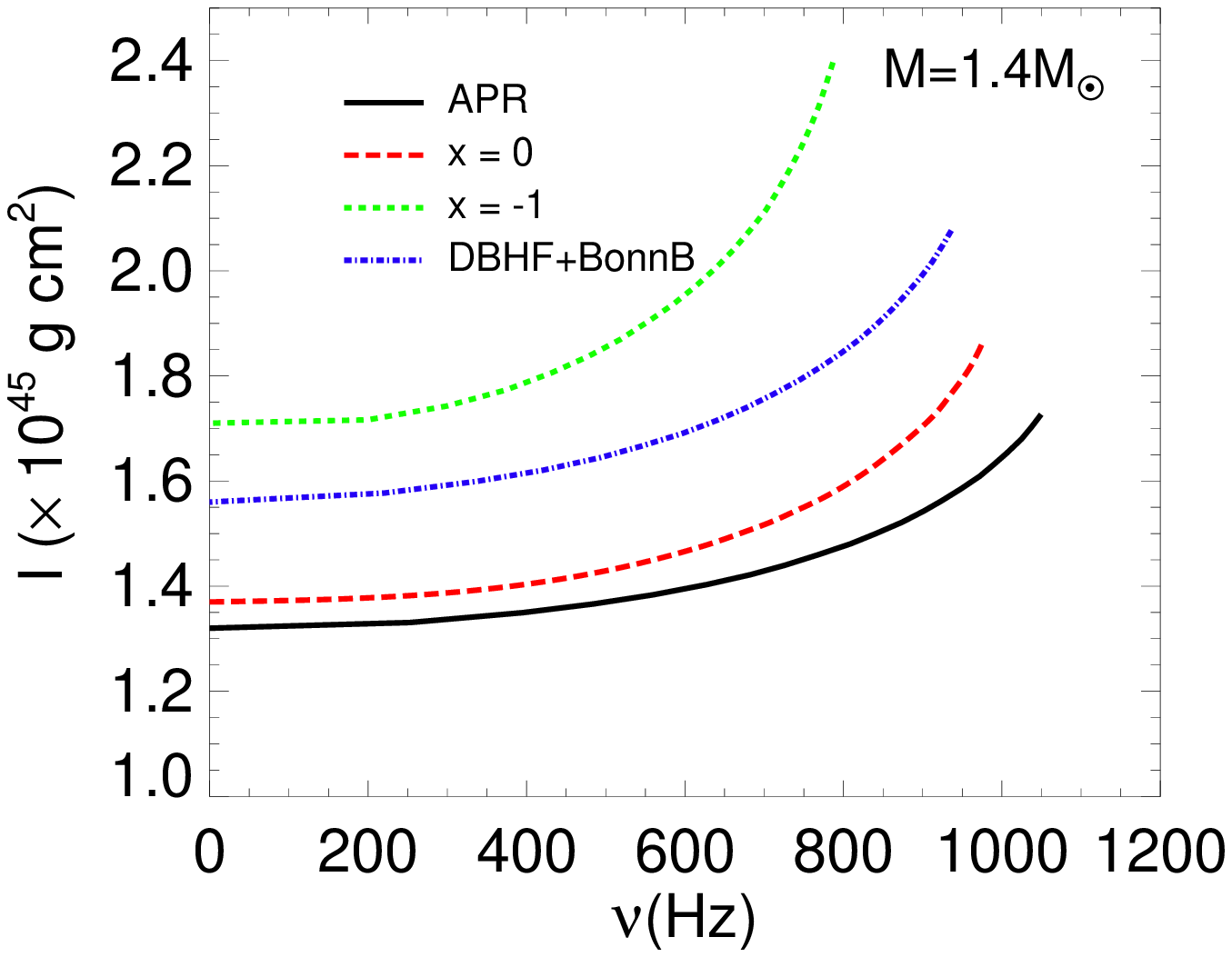}
\vspace{5mm} \caption{(Color online) Total moment of inertia as a
function of rotational frequency for stellar models with mass
$M=1.4M_{\odot}$. Taken from Ref. \cite{WKL:2008ApJ}.}\label{f15}
\end{figure}

In this subsection we turn our attention to the moment of inertia of
rapidly rotating neutron stars. In Fig.~\ref{f13} we show the moment
of inertia as a function of stellar mass for neutron star models
spinning at the mass-shedding (Kepler) frequency. The numerical
calculation is performed with the $RNS$ code. We observe that the
moments of inertia of rapidly rotating neutron stars are
significantly larger than those of slowly rotating models (for a
fixed mass). This is easily understood in terms of the increased
(equatorial) radius (Fig.~\ref{f5}).

We also compute the moments of inertia of pulsars spinning at 642Hz,
715Hz, and 1000Hz (642Hz \cite{Backer:1982} and 716Hz
\cite{Kaaret:2006gr} are the rotational frequencies of the fastest
pulsars as of today). The numerical results are presented in
Fig.~\ref{f14}. As demonstrated by Bejger et
al.~\cite{Bejger:2006hn} and most recently by Krastev et
al.~\cite{KLW2}, the range of the allowed masses supported by a
given EOS for rapidly rotating neutron stars becomes narrower than
the one for static configurations. The effect becomes stronger with
increasing frequency and depends upon the EOS. This is also
illustrated in Fig.~\ref{f14}, particularly in panel (c).
Additionally, the moment of inertia shows increase with rotational
frequency at a rate dependent upon the details of the EOS. This is
best seen in Fig.~\ref{f15} where we display the moment of inertia
as a function of the rotational frequency for stellar models with a
fixed mass ($M=1.4M_{\odot}$).

The neutron star sequences shown in Fig.~\ref{f14} are terminated at
the mass-shedding frequency. At the lowest frequencies the moment of
inertia remains roughly constant for all EOSs (which justifies the
application of the slow-rotation approximation and
Eq.~(\ref{eq.37})). As the stellar models approach the Kepler
frequency, the moment of inertia exhibits a sharp rise. This is
attributed to the large increase of the circumferential radius as
the star approaches the ``mass-shedding point''. As pointed by
Friedman et al.~\cite{1984Natur.312..255F}, properties of rapidly
rotating neutron stars display greater deviations from those of
spherically symmetric (static) stars for models computed with
stiffer EOSs. This is because such models are less centrally
condensed and gravitationally bound. This also explains why the
momenta of inertia of rapidly rotating neuron star configurations
from the $x=-1$ EOS show the greatest deviation from those of static
models.

\subsection{Fractional moment of inertia of the neutron star crust}

As it was discussed extensively by Lattimer and Prakash
\cite{Lattimer:2000kb} (and others), the neutron star crust
thickness might be measurable from observations of pulsar glitches,
the occasional disrupts of the otherwise extremely regular pulsation
from magnetized, rotating neutron stars. The canonical model of Link
et al. \cite{Link:1999} suggests that glitches are due to the
angular momentum transfer from superfluid neutrons to normal matter
in the neutron star crust, the region of the star containing nuclei
and nucleons that have dripped out of nuclei. This region is bound
by the neutron drip density at which nuclei merge into uniform
nucleonic matter. Link et al. \cite{Link:1999} concluded from the
observations of the Vela pulsar that at least $1.4\%$ of the total
moment of inertia resides in the crust of the Vela pulsar. For
slowly rotating neutron stars, applying several realistic hadronic
EOSs that permit maximum masses of at least $\sim 1.6M_{\odot}$
Lattimer and Prakash \cite{Lattimer:2000kb} found that the
fractional moment of inertia, $\Delta I/I$, can be expressed
approximately as
\begin{equation}\label{eq.38}
{\Delta I\over I}\simeq{28\pi P_t R^3\over
3Mc^2}{(1-1.67\beta-0.6\beta^2)\over\beta}\left[1+{2P_t(1+5\beta-14\beta^2)
\over \rho_tm_bc^2\beta^2}\right]^{-1}\
\end{equation}
In the above equation $\Delta I$ is the moment of inertia of the
neutron star crust, $I$ is the total moment of inertia, $\beta =
GM/Rc^2$ is the compactness parameter, $m_b$ is the average nucleon
mass, $\rho_t$ is the transition density at the crust-core boundary,
and $P_t$ is the transition pressure.
\begin{table}[t!]
\caption{Transition densities and pressures for the EOSs used in
this paper.}
\begin{center}
\begin{tabular}{lcccc}\label{tab.6}
EOS &  MDI(x=0) & MDI(x=-1) & APR & DBHF+Bonn B\\
\hline\hline
$\rho_t(fm^{-3})$     & 0.091 (0.095) & 0.093 (0.092) & 0.087 & 0.100\\
$P_t(MeV$ $fm^{-3})$  & 0.645 & 0.982 & 0.513 & 0.393\\
\hline
\end{tabular}
\end{center}
{\small The first row identifies the equation of state. The
remaining rows exhibit the following quantities: transition density,
transition pressure. The numbers in the parenthesis are the
transition densities calculated by Kubis \cite{Kubis:2007}.}
\end{table}
\begin{figure}[b!]
\centering
\includegraphics[totalheight=2.4in]{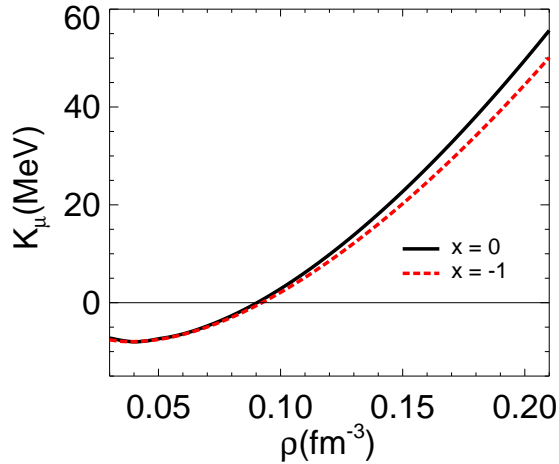}
\vspace{5mm} \caption{(Color online) The incompressibility,
$K_{\mu}$, as a function of baryon density $\rho$. Taken from
Ref.~\cite{WKL:2008ApJ}.}\label{f16}
\end{figure}
The determination of the transition density itself is a very
complicated problem. Different approaches often give quite different
results. Similar to determining the critical density for the
spinodal decomposition for the liquid-gas phase transition in
nuclear matter, for uniform $npe$-matter, Lattimer and Prakash
\cite{Lattimer:2000kb} and more recently Kubis \cite{Kubis:2007}
have evaluated the crust transition density by investigating when
the incompressibility of $npe$-matter becomes negative, i.e
\begin{equation}\label{eq.39}
K_{\mu}=\rho^2{d^2E_0\over d\rho^2}+2\rho{dE_0\over
d\rho}+\delta^2\left[\rho^2{d^2E_{sym}\over
d\rho^2}+2\rho{dE_{sym}\over
d\rho}-2E_{sym}^{-1}\left(\rho{dE_{sym}\over
d\rho}\right)^2\right]<0
\end{equation}
(see Fig.~\ref{f16}) where $E_0(\rho)$ is the EOS of symmetric
nuclear matter, $E_{sym}$ is the nuclear symmetry energy, and
$\delta=(\rho_n-\rho_p)/(\rho_n+\rho_p)$ is the asymmetry parameter.
Using this approach and the MDI interaction, Kubis \cite{Kubis:2007}
found the transition density of $0.119, 0.092, 0.095$ and $0.160
fm^{-3}$ for the $x$ parameter of $1, 0, -1 $ and $-2$,
respectively. Similarly, we have calculated the transition densities
and pressures for the EOSs employed in this work. Our results are
summarized in Table~\ref{tab.6}. We find good agreement between our
results and those by Kubis \cite{Kubis:2007} with the MDI
interaction. It is interesting to notice that the transition
densities predicted by all EOSs are in the same density range
explored by heavy-ion reactions at intermediate energies. The MDI
interaction with $x=0$ and $x=-1$ constrained by the available data
on isospin diffusion in heavy-ion reaction at intermediate energies
thus limits the transition density rather tightly in the range of
$\rho_t=[0.091-0.093](fm^{-3})$. It is worth noticing, however, that
the transition density is estimated here by using the parabolic
approximation of the EOS for isospin asymmetric nuclear matter.
Relaxing this approximation, using both dynamical and
thermodynamical approaches, the transition density and pressure have
been studied in great detail in Ref.~\cite{Xu09a}. A somewhat lower
values of $\rho_t=[0.040-0.065](fm^{-3})$ were obtained.
Nevertheless, to be consistent with the rest of the study in this
review, we stick to the parabolic approximation of the EOS in
studying the $\Delta I/I$.

\begin{figure}[t!]
\centering
\includegraphics[totalheight=2.4in]{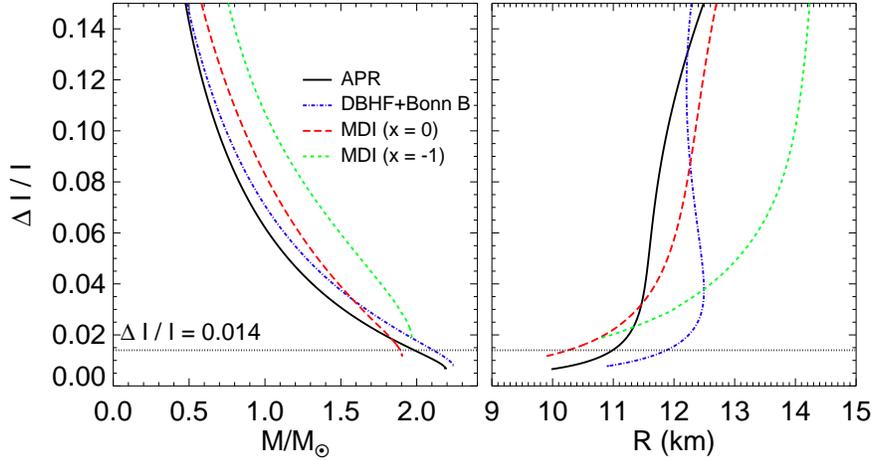}
\vspace{5mm} \caption{(Color online) The fractional moment of
inertia of the neutron star crust as a function of the neutron star
mass (left panel) and radius (right panel) estimated with
Eq.~(\ref{eq.39}). The constraint from the glitches of the Vela
pulsar is also shown. Taken from
Ref.~\cite{WKL:2008ApJ}.}\label{f17}
\end{figure}

The fractional momenta of inertia $\Delta I/I$ of the neutron star
crusts are shown in Fig.~\ref{f17} as computed through
Eq.~(\ref{eq.38}) with the parameters listed in Table~\ref{tab.6}.
It is seen that the condition $\Delta I/I>0.014$ extracted from
studying the glitches of the Vela pulsar does put a strict lower
limit on the radius for a given EOS. It also limits the maximum
mass to be less than about $2M_{\odot}$ for all of the EOSs
considered. Similar to the total momenta of inertia the ratio
$\Delta I/I$ changes more sensitively with the radius as the EOS
is varied.

\section{Rotation and proton fraction}

\begin{figure}[!t]
\centering
\includegraphics[totalheight=3.2in]{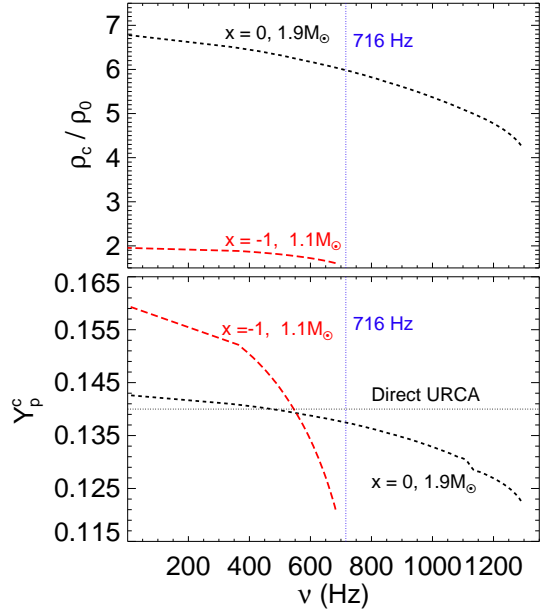}
\vspace{5mm} \caption{(Color online) Density (upper panel) and
proton fraction (lower panel) versus rotational frequency for fixed
neutron star mass. Taken from Ref. \cite{KLW2}.}\label{f18}
\end{figure}

\begin{figure}[!t]
\centering
\includegraphics[totalheight=3.2in]{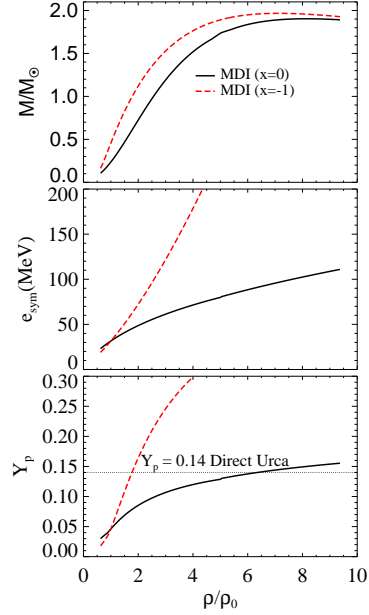}
\vspace{5mm} \caption{(Color online) Neutron star mass (upper
panel), nuclear symmetry energy (middle panel) and proton fraction
(lower panel) versus baryon density. Predictions from the MDI
(x=-1,0) are shown. Taken from Ref. \cite{KLW2}.}\label{f19}
\end{figure}

Finally, we study the effect of (fast) rotation on the proton
fraction in the neutron star core. In Fig.~\ref{f18} we show the
central baryon density (upper frame) and central proton fraction
(lower frame) as a function of the rotational frequency for
fixed-mass models. Predictions from both $x=0$ and $x=-1$ EOSs are
shown. We observe that central density decreases with increasing
frequency. This reduction is more pronounced in heavier neutron
stars. Most importantly, we also observe decrease in the proton
fraction $Y_p^c$ in the star's core. We recall that large proton
fraction (above $\sim 0.14$ for $npe\mu$-stars) leads to fast
cooling of neutron stars through direct Urca reactions. Our results
demonstrate that depending on the stellar mass and rotational
frequency, the central proton fraction could, in principle, drop
below the threshold for the direct {\it nucleonic} Urca channel and
thus making the fast cooling in rotating neutron stars impossible.
The masses of the models shown in Fig.~\ref{f18} are chosen so that
the proton fraction in stellar core is just above the direct Urca
limit for the {\it static} configurations, see Fig.~\ref{f19} upper
and lower frames. The stellar sequences in Fig.~\ref{f18} are
terminated at the Kepler (mass-shedding) frequency. In both cases
the central proton fraction drops below the direct Urca limit at
frequencies lower than that of PSR J1748-244ad
\cite{Hessels:2006ze}. This implies that the fast cooling can be
effectively blocked in millisecond pulsars depending on the exact
mass and spin rate. It might also explain why heavy neutron stars
(could) exhibit slow instead of fast cooling. For instance, with the
$x=0$ EOS (with softer symmetry energy) for a neutron star of mass
approximately $1.9M_{\odot}$, the Direct Urca channel closes at
$\nu\approx 470Hz$. On the other hand, with the x=-1 EOS (with
stiffer symmetry energy) the direct Urca channel can close only for
low mass neutron stars, in fact only for masses well below the
canonical mass of $1.4M_{\odot}$. This is due to the much stiffer
symmetry energy (see Fig.~\ref{f19} middle frame) because of which
the direct Urca threshold (Fig.~\ref{f19} lower frame) is reached at
much lower densities and stellar masses (Fig.~\ref{f19} upper
frame).

Before closing the discussion in this section a few comments are in
order. In the present study we do not consider ''exotic" states of
matter in neutron stars. On the other hand, due to the rapid rise of
the baryon chemical potentials several other species of particles,
such as strange hyperons $\Lambda^0$ and $\Sigma^-$, are expected to
appear once their mass thresholds are
reached~\cite{2000PhRvC..61e5801B}. The appearance of hyperonic
degrees of freedom lowers the energy-per-particle in the stellar
medium and causes more centrally condensed configurations with lower
masses and radii. Additionally, hyperons help the condition for {\it
nucleonic} direct Urca process to be satisfied at lower densities
due to the increased proton fraction \cite{1992ApJ...390L..77P}, and
depending on their exact concentrations could potentially contribute
to the fast cooling of the star through {\it hyperonic} direct Urca
processes \cite{Page:2005fq}. These considerations would alter the
balance between the curves in Fig.~\ref{f19} and ultimately the
results displaced in Fig.~\ref{f18} in favor of the direct Urca
process, i.e. smaller masses and higher frequencies would be
necessary to close the fast cooling channel. This is due to the fact
that the overall impact of (rapid) rotation on the neutron star
structure is smaller for more centrally condensed models resulting
from ``softer'' EOSs \cite{1984Natur.312..255F}. Therefore, for such
models there is smaller deviation from properties and structure of
static configurations. In addition, at even higher densities matter
is expected to undergo a transition to quark-gluon plasma
\cite{Weber:1999a,2000PhRvC..61e5801B}, which favors a fast cooling
through enhanced nucleonic direct Urca and quark direct Urca
processes, see e.g., Ref. \cite{Page:2005fq}.

\section{Constraining gravitational waves from elliptically deformed pulsars}

Gravitational waves are tiny disturbances in space-time and are a
fundamental, although not yet directly confirmed, prediction of
General Relativity. They can be triggered in cataclysmic events
involving (compact) stars and/or black holes. They could even have
been produced during the very early Universe, well before any stars
had been formed, merely as a consequence of the dynamics and
expansion of the Universe. Because gravity interact extremely weakly
with matter, gravitational waves would carry a genuine picture of
their sources and thus provide undisturbed information that no other
messenger can deliver~\cite{Maggiore:2007}. Gravitational wave
astrophysics would open an entirely new non-electromagnetic window
making it possible to probe physics that is hidden or dark to
current electromagnetic observations~\cite{Flanagan:2005yc}.

(Rapidly) rotating neutron stars could be one of the major
candidates for sources of continuous gravitational waves in the
frequency bandwidth of the LIGO~\cite{Abbott:2004ig} and VIRGO
(e.g. Ref.~\cite{Acernese:2007zzb}) laser interferometric
detectors. It is well known that a rotating object self-bound by
gravity and which is perfectly symmetric about the axis of
rotation does not emit gravitational waves. In order to generate
gravitational radiation over extended period of time, a rotating
neutron star must have some kind of long-living axial
asymmetry~\cite{Jaranowski:1998qm}. Several mechanisms leading to
such an asymmetry have been studied in the literature: (1) Since
the neutron star crust is solid, its shape might not be
necessarily symmetric, as it would be for a fluid, with
asymmetries supported by anisotropic stress built up during the
crystallization period of the crust \cite{PPS:1976ApJ}. (2)
Additionally, due to its violent formation (supernova) or due to
its environment (accretion disc), the rotational axis may not
coincide with a principal axis of the moment of inertia of the
neutron star which make the star precess~\cite{ZS:1979PRD}. Even
if the star remains perfectly symmetric about the rotational axis,
since it precesses, it emits gravitational waves
\cite{ZS:1979PRD,Z:1980PRD}. (3) Also, the extreme magnetic fields
presented in a neutron star cause magnetic pressure (Lorenz forces
exerted on conducting matter) which can distort the star if the
magnetic axis is not aligned with the axis of
rotation~\cite{BG:1996AA}, which is widely supposed to occur in
order to explain the pulsar phenomenon. Several other mechanisms
exist that can produce gravitational waves from neutron stars. For
instance, accretion of matter on a neutron star can drive it into
a non-axisymmetric configuration and power steady radiation with a
considerable amplitude~\cite{W:1984ApJ}. This mechanism applies to
a certain class of neutron stars, including accreting stars in
binary systems that have been spun up to the first instability
point of the so-called Chadrasekhar-Friedman-Schutz (CFS)
instability~\cite{Schutz:1997}. Also,
Andersson~\cite{Andersson:1997xt} suggested a similar instability
in $r$-modes of (rapidly) rotating relativistic stars. It has been
shown that the effectiveness of these instabilities depends on the
viscosity of stellar matter which in turn is determined by the
star's temperature.

Gravitational waves are characterized by tiny dimensionless strain
amplitude, which depends on the degree to which the neutron star is
deformed from axial symmetry which, in turn, depends upon the
equation of state (EOS) of neutron-rich stellar matter. As already
mentioned, presently the EOS of matter under extreme conditions
(densities, pressures and isospin asymmetries) is still rather
uncertain, mainly due to the poorly known density dependence of the
nuclear symmetry energy, $E_{sym}(\rho)$,
e.g.~\cite{Lattimer:2004pg}. Applying several nucleonic EOSs
(discussed in section 2), we calculate the gravitational wave strain
amplitude for selected neutron star configurations. Particular
attention is paid to predictions with the EOS partially constrained
by the nuclear laboratory data. These results set an upper limit on
the strain amplitude of gravitational radiation expected from
rotating neutron stars.

The pulsar population is such that most have spin frequencies that
fall below the sensitivity band of current detectors. In the future,
the low-frequency sensitivity of VIRGO~\cite{Acernese:2005} and
Advanced LIGO~\cite{Creighton:2003} should allow studies of a
significantly larger sample of pulsars. Moreover, LISA  (the Laser
Interferometric Space Antenna) is currently being jointly designed
by NASA in the United States and ESA (the European Space Agency),
and will be launched into orbit in the near future providing an
unprecedented instrument for gravitational waves search and
detection~\cite{Flanagan:2005yc}. The discussion in this section is
(mainly) based on Ref. \cite{Krastev:2008PLB}.

\subsection{Formalism relating the EOS of neutron-rich matter to the
strength of gravitational waves from slowly rotating neutron stars}

In the following we review briefly the formalism used to calculate
the gravitational wave strain amplitude from slowly rotating neutron
stars. A spinning neutron star is expected to emit GWs if it is not
perfectly symmetric about its rotational axis. As already mentioned,
non-axial asymmetries can be achieved through several mechanisms
such as elastic deformations of the solid crust or core or
distortion of the whole star by extremely strong misaligned magnetic
fields. Such processes generally result in a triaxial neutron star
configuration~\cite{Abbott:2004ig} which, in the quadrupole
approximation and with rotation and angular momentum axes aligned,
would cause gravitational waves at {\it twice} the star's rotational
frequency~\cite{Abbott:2004ig}. These waves have characteristic
strain amplitude at the Earth's vicinity (assuming an optimal
orientation of the rotation axis with respect to the observer)
of~\cite{HAJS:2007PRL}
\begin{equation}\label{eq.40}
h_0=\frac{16\pi^2G}{c^4}\frac{\epsilon I_{zz}\nu^2}{r},
\end{equation}
where $\nu$ is the neutron star rotational frequency, $I_{zz}$ its
principal moment of inertia, $\epsilon=(I_{xx}-I_{yy})/I_{zz} $ its
equatorial ellipticity, and $r$ its distance to Earth. The
ellipticity is related to the neutron star maximum quadrupole moment
(with $m=2$) via~\cite{Owen:2005PRL}
\begin{equation}\label{eq.41}
\epsilon = \sqrt{\frac{8\pi}{15}}\frac{\Phi_{22}}{I_{zz}},
\end{equation}
where for {\it slowly} rotating (and static) neutron stars
$\Phi_{22}$ can be written as~\cite{Owen:2005PRL}
\begin{equation}\label{eq.42}
\Phi_{22,max}=2.4\times
10^{38}g\hspace{1mm}cm^2\left(\frac{\sigma}{10^{-2}}\right)\left(\frac{R}{10km}\right)^{6.26}
\left(\frac{1.4M_{\odot}}{M}\right)^{1.2}
\end{equation}
In the above expression $\sigma$ is the breaking strain of the
neutron star crust which is rather uncertain at present time.
Earlier studies have estimated $\sigma$ to be in the range
$\sigma=[10^{-5}-10^{-2}]$~\cite{HAJS:2007PRL}. More recently, using
molecular dynamics simulations, it was estimated to be about $0.1$
which is considerably larger than the previous
findings~\cite{Hor09}. In our work~\cite{Krastev:2008PLB}, we have
used $\sigma=10^{-2}$, which, compared to the latest
estimate~\cite{Hor09}, is rather conservative. Nevertheless, our
results are simply amplified by a factor of 100 if we apply the
estimate by Horowitz et al.~\cite{Hor09} in our calculations.  From
Eqs.~(\ref{eq.40}) and (\ref{eq.41}) it is clear that $h_0$ does not
depend on the moment of inertia $I_{zz}$, and that the total
dependence upon the EOS is carried by the quadrupole moment
$\Phi_{22}$. Thus Eq.~(\ref{eq.40}) can be rewritten as
\begin{equation}\label{eq.43}
h_0=\chi\frac{\Phi_{22}\nu^2}{r},
\end{equation}
with $\chi=\sqrt{2045\pi^5/15}G/c^4$. Eq. (\ref{eq.43}) establishes
the link between the gravitational wave strain amplitude and the EOS
of neutron-rich matter--the EOS is the main ingredient for
determining the neutron star properties, including its quadrupole
moment. In a recent work~\cite{WKL:2008ApJ} we have calculated the
neutron star moment of inertia of both static and (rapidly) rotating
neutron stars. For slowly rotating neutron stars Lattimer and
Schutz~\cite{Lattimer:2005} derived an empirical relation (Eq.
(\ref{eq.37})) for $I_{zz}$ which is shown to hold for a wide class
of equations of state which do not exhibit considerable softening
and for neutron star models with masses above
$1M_{\odot}$~\cite{Lattimer:2005}. Using Eq.~(\ref{eq.37}) to
calculate the neutron star moment of inertia and Eq.~(\ref{eq.42})
the corresponding quadrupole moment, the ellipticity $\epsilon$ can
be readily computed (via Eq.~(\ref{eq.41})). Since the global
properties of spinning neutron stars (in particular the moment of
inertia) remain approximately constant for rotating configurations
at frequencies up to $\sim 300Hz$~\cite{WKL:2008ApJ}, the above
formalism can be readily employed to estimate the gravitational wave
strain amplitude, provided one knows the exact rotational frequency
and distance to Earth, and that the frequency is relatively low
(below $\sim 300Hz$). These estimates are then to be compared with
the current upper limits for the sensitivity of the laser
interferometric observatories (e.g. LIGO).

\subsection{Gravitational waves from {\it slowly} rotating neutron stars}

We calculate the gravitational wave strain amplitude $h_0$ for
several selected pulsars with rotational frequencies $\sim 200Hz$
and relatively close to Earth, employing several nucleonic equations
of state. We assume a simple model of stellar matter of nucleons and
light leptons (electrons and muons) in beta-equilibrium. The
equations of state that we employ here have been discussed in
section 2 and shown in Fig. \ref{f3}.

\begin{figure}[t!]
\centering
\includegraphics[height=5.5cm]{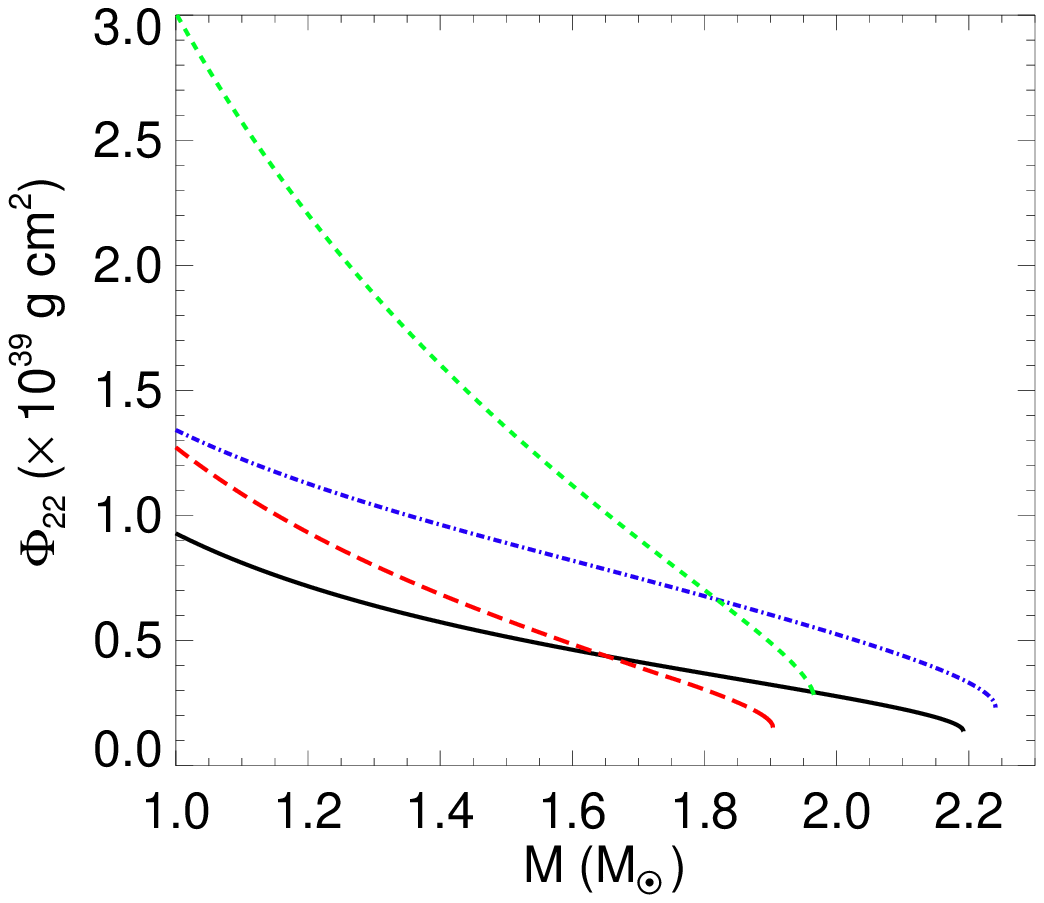}
\includegraphics[height=5.5cm]{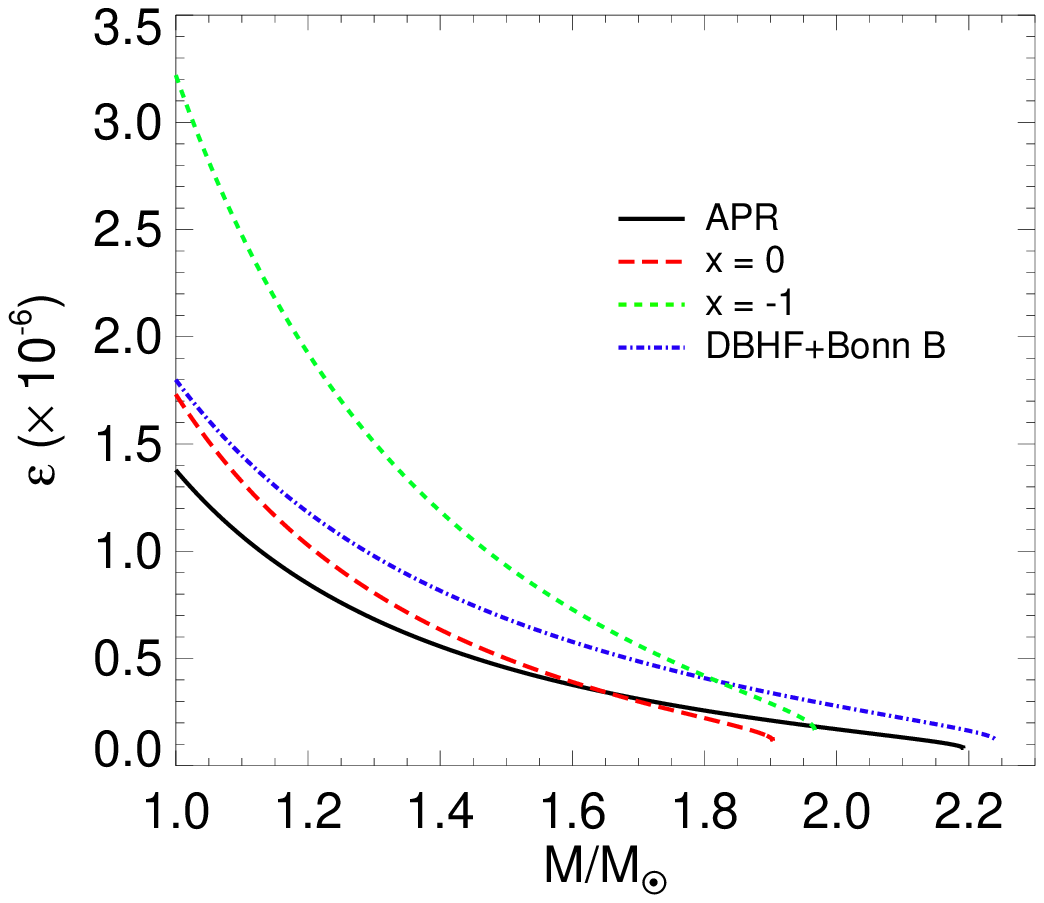}
\caption{(Color online) Neutron star quadrupole moment (left panel)
and ellipticity (right panel). Taken from Ref.
\cite{Krastev:2008PLB}.} \label{f20}
\end{figure}

Fig.~\ref{f20} displays the neutron star quadrupole moment (left
panel) and ellipticity (right panel). The quadrupole moment is
calculated through Eq.~(\ref{eq.42}) and the ellipticity through Eq.
(\ref{eq.41}). Note that Eq.~(\ref{eq.42}) is valid only for {\it
slowly} rotating neutron star models. We notice that $\Phi_{22}$
decreases with increasing stellar mass for all EOSs considered in
this study. The rate of this decrease depends upon the EOS and is
largest for the $x=-1$ EOS. This behavior is easily understood in
terms of the increased central density with stellar mass -- more
massive stars are more compact and, since the quadrupole moment is a
measure of the star's deformation (see Eq.~(\ref{eq.41})), they are
also less deformed with respect to less centrally condensed models.
Moreover, it is well known that the mass is mainly determined by the
symmetric part of the EOS while the radius of a neutron star is
strongly affected by the density slope of the symmetry energy. More
quantitatively, an EOS with a stiffer symmetry energy, such as the
$x=-1$ EOS, results in less compact stellar models, and hence more
deformed pulsars. Here we recall specifically that the $x=-1$ EOS
yields neutron star configurations with larger radii than those of
models form the rest of the EOSs considered in this study (e.g. see
Fig.~\ref{f5}). These results are consistent with previous findings
which suggest that more compact neutron star models are less altered
by rotation, e.g. see Ref.~\cite{1984Natur.312..255F}. Consequently,
it is reasonable also to expect such configurations to be more
``resistant'' to any kind of deformation. The neutron star
ellipticity, $\epsilon$, is shown as a function of the neutron star
mass (Fig.~\ref{f20}, right panel). Since $\epsilon$ is proportional
to the quadrupole moment $\Phi_{22}$ (scaled by the moment of
inertia $I_{zz}$), it decreases with increasing stellar mass. The
results shown in the right panel of Fig.~\ref{f20} are consistent
with the maximum ellipticity $\epsilon_{max}\approx 2.4\times
10^{-6}$ corresponding to the largest crust ``mountain'' one could
expect on a neutron star~\cite{HAJS:2007PRL,HJA:2006MNRAS}. (The
estimate of $\epsilon_{max}$ in
Refs.~\cite{HAJS:2007PRL,HJA:2006MNRAS} has been obtained assuming
breaking strain of the crust $\sigma=10^{-2}$, as we have assumed in
the present paper in calculating the neutron star quadrupole
moment.)

\begin{figure}[!t]
\centering
\includegraphics[totalheight=5.5cm]{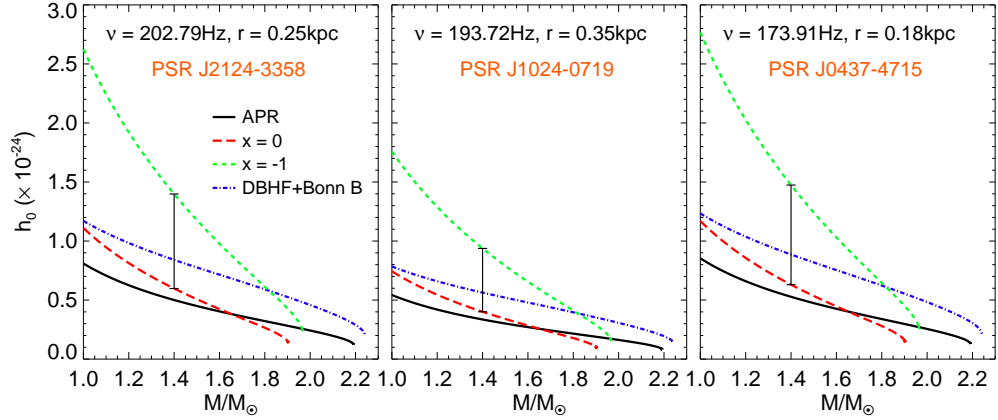}
\vspace{5mm} \caption{(Color online) Gravitational-wave strain
amplitude as a function of the neutron star mass. The error bars
between the $x=0$ and $x=-1$ EOSs provide a limit on the strain
amplitude of the gravitational waves to be expected from these
neutron stars, and show a specific case for stellar models of
$1.4M_{\odot}$. Taken form Ref.~\cite{Krastev:2008PLB}.} \label{f21}
\end{figure}

In Fig.~\ref{f21} we display the GW strain amplitude, $h_0$, as a
function of stellar mass. Predictions are shown for three selected
millisecond pulsars, which are relatively close to Earth
($r<0.4kpc$), and have rotational frequencies below $300Hz$ so that
the corresponding momenta of inertia and quadrupole moments can be
computed approximately via Eqs.~(\ref{eq.37}) and (\ref{eq.42})
respectively. The properties of these pulsars (of interest to this
study) are summarized in Table~\ref{tab.7}.
\begin{table}[!t]
\begin{center}
\caption{Properties of the pulsars considered in this
study.}\vspace{3mm}
\begin{tabular}{lcccc}\label{tab.7}
Pulsar        & $\nu(Hz)$ & $M(M_{\odot})$ & $r(kpc)$   & Reference\\
\hline\hline
PSR J2124-3358 & 202.79   & -            &  0.25      &  \cite{Biales:1997ApJ}                   \\
PSR J1024-0719 & 193.72   & -            &  0.35      &  \cite{Biales:1997ApJ}                   \\
PSR J0437-4715 & 173.91   & $1.3\pm 0.2$ &  0.18      &  \cite{HBO:2006MNRAS,Straten:2001Nature} \\
\hline
\end{tabular}
\end{center}
\vspace{3mm} {\small The first column identifies the pulsar. The
remaining columns exhibit the following quantities: rotational
frequency; mass (if known); distance to Earth; corresponding
references. Notice that only the mass of PSR J0437-4715 is known
from orbital dynamics~\cite{Straten:2001Nature,HBO:2006MNRAS} (as
the pulsar has a low-mass white dwarf companion). The masses of PSRs
J2124-3358 and J1024-0719 are presently unknown as they are both
isolated neuron stars~\cite{Biales:1997ApJ}.}
\end{table}
The error bars in Fig.~\ref{f21} between the $x=0$ and $x=-1$ EOSs
provide a constraint on the {\it maximal} strain amplitude of the
gravitational waves emitted by the millisecond pulsars considered
here. The specific case shown in the figure is for neutron star
models of $1.4M_{\odot}$. Depending on the exact rotational
frequency, distance to detector, and details of the EOS, the {\it
maximal} $h_0$ is in the range $\sim[0.4-1.5]\times 10^{-24}$. These
estimates do not take into account the uncertainties in the distance
measurements. They also should be regarded as upper limits since the
quadrupole moment (Eq.~(\ref{eq.42})) has been calculated with
$\sigma=10^{-2}$ (where $\sigma$ can go as low as $10^{-5}$). Here
we recall that the mass of PSR J0437-4715 (Fig.~\ref{f21} right
panel) is $1.3\pm 0.2M_{\odot}$ \cite{HBO:2006MNRAS}. (Another mass
constraint, $1.58 \pm 0.18M_{\odot}$, was given previously by van
Straten et al. \cite{Straten:2001Nature}.) The results shown in
Fig.~\ref{f21} suggest that the GW strain amplitude depends on the
EOS of stellar matter, where this dependence is stronger for lighter
neutron star models. In addition, it is also greater for stellar
configurations computed with stiffer EOS. As explained, such models
are less compact and thus less gravitationally bound. As a result,
they could be more easily deformed by rotation or/and other
deformation driving mechanisms and phenomena, and therefore are
expected to emit stronger gravitational radiation
(Eq.~(\ref{eq.40})).

\begin{figure}[!t]
\centering
\includegraphics[totalheight=6.0cm]{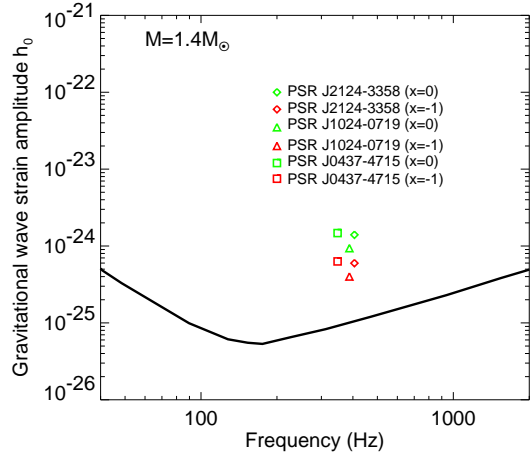}
\vspace{5mm} \caption{(Color online) Gravitational wave strain
amplitude as a function of the gravitational wave frequency. The
characters denote the strain amplitude of the GWs expected to be
emitted from spinning neutron stars ($\nu<300Hz$) with mass
$1.4M_{\odot}$. Solid line denotes the current upper limit of the
LIGO sensitivity. Adapted from Ref.~\cite{Abbott:2004ig}.}
\label{f22}
\end{figure}

In Fig.~\ref{f22} we take another view of the results shown in
Fig.~\ref{f21}. We display the maximal GW strain amplitude as a
function of the GW frequency and compare our predictions with the
best current detection limit of LIGO. The specific case shown is for
neutron star models with mass $1.4M_{\odot}$ computed with the $x=0$
and $x=-1$ EOSs. Since these EOSs are constrained by the available
nuclear laboratory data they provide a limit on the possible neutron
star configurations and thus gravitational emission from them. The
results shown in Fig.~\ref{f22} would suggest that presently the
gravitational radiation from the three selected pulsars should be
within the detection capabilities of LIGO. The fact that such a
detection has not been made yet deserves a few comments at this
point. First, as we mentioned, in the present calculation we assume
breaking strain of the neutron star crust $\sigma=10^{-2}$ which
might be too optimistic. As pointed by Haskell et
al.~\cite{HAJS:2007PRL}, we are still away from testing more
conservative models and if the true value of $\sigma$ lies in the
low end of its range, we would be still far away from a direct
detection of a gravitational wave signal. Second, while we have
assumed a specific neutron star mass of $1.4M_{\odot}$,
Fig.~\ref{f21} tells us that $h_0$ decreases with increasing stellar
mass, i.e. heavier neutron stars will emit weaker GWs. Here we
recall that from the selected pulsars only the mass of PSR
J0437-4715 is known (within some
accuracy~\cite{HBO:2006MNRAS,Straten:2001Nature}). The masses of PSR
J2124-3358 and PSR J1024-0719 are unknown. Third, in the present
study we assume a very simple model of stellar matter consisting
only beta equilibrated nucleons and light leptons (electrons and
muons). On the other hand, in the core of neutron stars conditions
are such that other more exotic species of particles could readily
abound. Such novel phases of matter would soften considerably the
EOS of stellar medium \cite{2000PhRvC..61e5801B} leading to
ultimately more compact and gravitationally tightly bound objects
which could withstand larger deformation forces (and torques).
Lastly, the existence of quark stars, truly exotic self-bound
compact objects, is not excluded from further considerations and
studies. Such stars would be able to resist huge forces (such as
those resulting from extremely rapid rotation beyond the Kepler, or
mass-shedding, frequency) and as a result retain their axial
symmetric shapes effectively dumping the gravitational radiation
(e.g. Ref. \cite{Weber:1999a}). At the end, we recall that
Eq.~(\ref{eq.40}) implies that the best possible candidates for
gravitational radiation (from spinning relativistic stars) are {\it
rapidly} rotating pulsars relatively close to Earth ($h_0\sim
\Phi_{22}\nu^2/r$). Increasing rotational frequency (and/or
decreasing distance to detector, $r$) would alter the results shown
in Figs.~\ref{f21} and \ref{f22} in favor of a detectable signal by
the current observational facilities (e.g. LIGO). On the other hand,
for more realistic and quantitative calculations, the neutron star
quadrupole moment must be calculated numerically exactly by solving
the Einstein field equations for rapidly rotating neutron stars.
(Such calculations have been reported, for instance, by Laarakkers
and Piosson~\cite{LP:1999ApJ}.)

\subsection{Gravitational waves from {\it rapidly} rotating neutron stars}

For rapidly rotating neutron stars the estimate of the quadrupole
deformation in Eq. (\ref{eq.42}) is no longer valid. Moreover, the
moment of inertia has to be calculated differently as well. Often,
one uses the spin-down rate to estimate the strain amplitude for
fast pulsars. Provided that the spin-down rate, $\dot{\nu}$, for a
given pulsar is known, $h_0$ could be estimated from
\cite{Abbott:PRD2007}
\begin{equation}\label{eq.44}
h_0^{sd}=\frac{5}{2}\left(\frac{GI_{zz}|\dot{\nu}|}{c^3r^2\nu}\right)^{1/2}.
\end{equation}
Here we should mention that in such calculations one assumes that
the {\it only} mechanism contributing to the pulsar's observed
spin-down is gravitational radiation. However, other mechanisms
could also account for the star's observed decrease in rotational
frequency such as magnetic dipole radiation, and particle
acceleration in the magnetosphere~\cite{Abbott:2008APL}. Despite
these uncertainties, calculations of gravitational wave strain
amplitude through Eq.~(\ref{eq.44}) are still important because, in
addition to providing a rather conservative upper limit on the
expected gravitational radiation, they also serve to estimate
another very uncertain but important quantity -- the ellipticity
$\epsilon$. So far the fastest pulsars reported are the PSR
B1937+21~\cite{Backer:1982} and PSR
J1748-2446ad~\cite{Hessels:2006ze}. Their known properties are
summarized in Table \ref{tab.8}.
\begin{table}[!t]
\caption{Properties of the {\it rapidly rotating} pulsars considered
in this study. The first column identifies the pulsar. The remaining
columns exhibit the following quantities: rotational frequency;
first derivative of the rotational frequency; distance to Earth;
corresponding reference.}\vspace{2mm}\centering
\begin{tabular}{lcccc}\label{tab.8}
Pulsar        & $\nu(Hz)$ & $\dot{\nu}(Hz$ $s^{-1})$ & $r(kpc)$ &Reference  \\
\hline\hline
PSR B1937+21      & 641.93   & $-$4.33$\times$ $10^{-14}$  &  3.60  &  \cite{Backer:1982,Cusmano:2004NPPS}     \\
PSR J1748-2446ad  & 716.35   & --                          &  8.70  &  \cite{Hessels:2006ze,Manchester:2005AJ} \\
\end{tabular}
\end{table}

The $RNS$ code~\cite{Stergioulas:1994ea} was used to calculate the
principal moment of inertia (and other properties) of rapidly
rotating neutron stars. The neutron star moment of inertia for the
two fastest pulsars is shown in Fig \ref{f14}. Panel (a) shows the
moment of inertia of PSR B1937+21 and panel (b) displays the moment
of inertia of PSR J1748-2446ad. As already observed previously, the
moment of inertia increases with rotational frequency, while the
range of possible neutron star configurations decreases (see, for
instance, Refs.~\cite{KLW2,WKL:2008ApJ}).

\begin{figure}[t!]
\centering
\includegraphics[height=5cm]{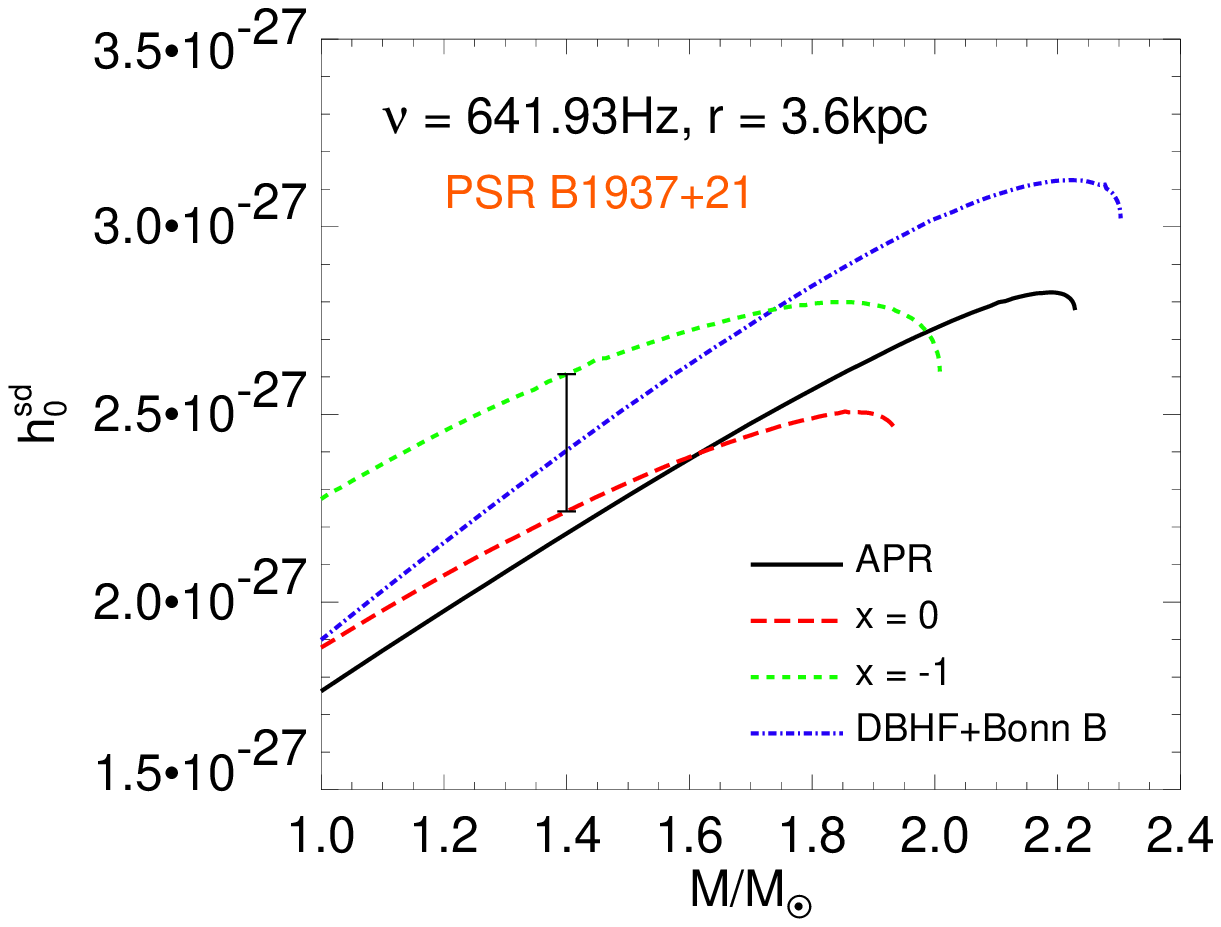}
\includegraphics[height=5cm]{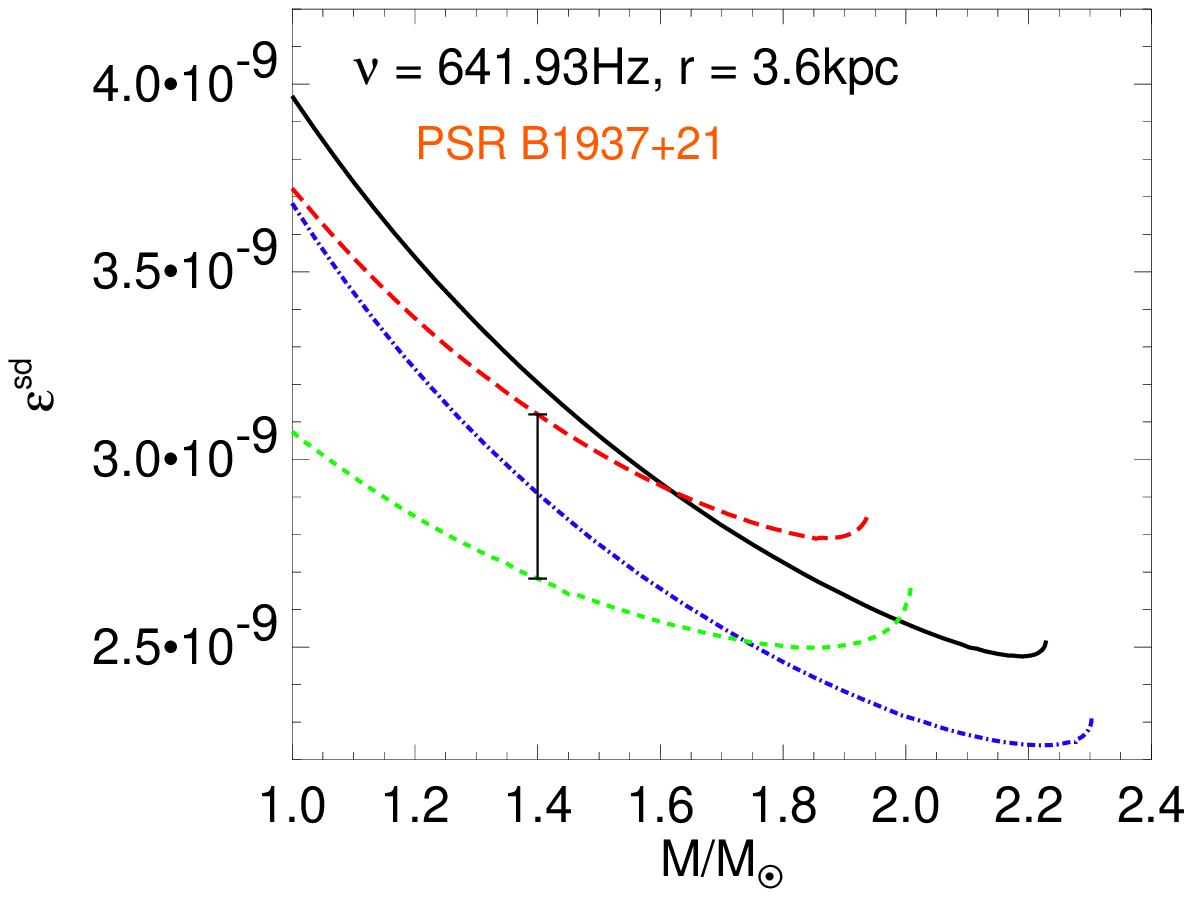}
\caption{(Color online) Gravitational wave strain amplitude
$h_0^{sd}$ (left panel) and ellipticity (right panel), deduced from
the spin-down rate of PSR B1937+21. See text for details.}
\label{f23}
\end{figure}

A particularly interesting example is the neutron star rotating at
642 Hz~\cite{Backer:1982}. Since its first observation in 1982, this
pulsar has been studied extensively and an observed spin-down rate
has been measured (see Table 1). Using the spin-down rate, an upper
limit on the gravitational wave strain amplitude was obtained. The
spin-down rate corresponds to a loss in kinetic energy at a rate of
$\dot{E}$ = $4\pi^2I_{zz}\nu|\dot{\nu}|\sim$[0.6 -- 3.1] $\times$
10$^{36}erg/s$, depending on the EOS. Assuming that the energy loss
is completely due to the gravitational radiation, the gravitational
wave strain amplitude can be calculated through Eq.~(\ref{eq.44}).
Similar calculations for this pulsar and others with an observed
spin-down rate have been performed in the
past~\cite{Abbott:2004ig,Abbott:PRD2007}. These calculations have
provided estimates for the gravitational wave strain amplitude of
selected pulsars for which the spin-down rates are known, and also
upper bounds for their ellipticities using the quadrupole
model~\cite{Abbott:2004ig}. However, such calculations simply used
the ``fiducial'' value of $10^{45}$$g$ $cm^2$ for the moment of
inertia $I_{zz}$ in all estimates. On the other hand, the neutron
star moment of inertia is sensitive to the details of the EOS of
stellar matter, and especially to the density dependence of the
nuclear symmetry energy~\cite{WKL:2008ApJ}. Moreover, $I_{zz}$
increases with increasing rotational frequency (see, e.g.
Fig.~\ref{f15}) and the differences with the static values of the
moment of inertia could be significant, particularly for rapidly
rotating neutron stars.

The gravitational wave strain amplitude of PSR B1937+21 is shown in
Fig.~\ref{f23} (left panel). Because the MDI EOS is constrained by
available nuclear laboratory data, our results with the $x=0$ and
$x=-1$ EOSs allowed us to place a rather conservative {\it upper}
limit on the gravitational waves to be expected from this pulsar,
provided the {\it only} mechanism accounting for its spin-down rate
is gravitational radiation. Under these circumstances, the upper
limit of the strain amplitude, $h_0^{sd}$, for neutron star models
of $1.4M_{\odot}$ is in the range $h_0^{sd}=[2.24-2.61]\times
10^{-27}$. Similarly, we have constrained the upper limit of the
ellipticity of PSR B1937+21 to be in the range
$\epsilon^{sd}=[2.68-3.12]\times 10^{-9}$ (Fig.~\ref{f23}, right
panel).

\section{Summary and outlook}

In this chapter we have summarized our recent studies on
properties of (rapidly) rotating neutron stars and the
gravitational waves expected from deformed pulsars using several
nuclear EOSs constrained partially by terrestrial laboratory
experiments. In particular, the latest heavy-ion reaction
experiments have constrained partially the density dependence of
the nuclear symmetry energy and thus the EOS of neutron-rich
nuclear matter. These limits, while being incomplete and still
suffer from some remaining uncertainties, can already provide some
useful information about the possible stable configurations of
(rapidly) rotating neutron stars and the gravitational radiation
expected from them. Specifically, we have studied properties of
(rapidly) rotating neutron stars employing four nucleonic EOSs and
find that the rapid rotation affects the neutron star structure
significantly. It increases the maximum possible mass up to $\sim
17\%$ and increases/decreases the equatorial/polar radius by
several kilometers. The neutron star moment of inertia has been
studied for both slowly and rapidly rotating models within a well
established formalism. We found that the moment of inertia of PSR
J0737-3039A is limited in the range of $I=(1.30-1.63)\times
10^{45}(g$ $cm^2)$. The fractional momenta of inertia $\Delta I/I$
of the neutron star crust are also constrained. It is also found
that the moment of inertia increases with rotational frequency at
a rate strongly depending upon the EOS used. Additionally,
rotation reduces central density and proton fraction in the
neutron star core, and depending on the exact stellar mass and
rotational frequency could effectively close the fast cooling
channel in millisecond pulsars. This circumstance may have
important consequences for both the interpretation of cooling data
and the thermal evolution modelling.

We have reported predictions on the upper limit of the strain
amplitude of the gravitational waves to be expected from
elliptically deformed pulsars at frequencies $<300Hz$. Our results
are intended to provide guidance to the ground-based gravitational
wave observatories. By applying an EOS with symmetry energy
constrained by recent nuclear laboratory data, we obtained an upper
limit on the gravitational-wave signal to be expected from several
pulsars. Depending on details of the EOS, for several millisecond
pulsars $0.18kpc$ to $0.35Kpc$ from Earth, the {\it maximal} $h_0$
is found to be in the range of $\sim[0.4-1.5]\times 10^{-24}$.
Finally, from the spin-down rate of PSR B1937+21 we have deduced the
upper limit of the strain amplitude, $h_0^{sd}$, for neutron star
models of $1.4M_{\odot}$ to be in the range
$h_0^{sd}=[2.24-2.61]\times 10^{-27}$. We have also constrained the
upper limit of the ellipticity of PSR B1937+21 to be in the range
$\epsilon^{sd}=[2.68-3.12]\times 10^{-9}$.  These predictions set
the first direct nuclear constraints on the gravitational waves from
elliptically deformed pulsars.

Looking forward, new experiments in terrestrial nuclear laboratories
are expected to improve our understanding about the EOS of dense
neutron-rich nuclear matter dramatically in the next few years. The
EOS of neutron-rich matter will be better constrained in a wide
density range. Most of the studies presented in this review will
then be further refined. Conversely, rapid progress in astrophysical
observations of neutron stars and gravitational waves surely will
also help us better understand the EOS of neutron-rich matter, and
thus stimulate the progress in nuclear physics. Eventually, the
ultimate goal of understanding thoroughly all mysteries of pulsars
can only be realized by utilizing the progress made in both nuclear
physics and astrophysics.

\section*{Acknowledgements}
We thank Aaron Worley for his collaboration on some of the work
reported here. This work is supported in part by the US National
Science Foundation under Grants No. PHY0757839, the Research
Corporation under Award No. 7123 and the Texas Coordinating Board
of Higher Education Grant No.003565-0004-2007.

\label{lastpage-01}

\end{document}